\documentclass[3p]{elsarticle}

\let\today\relax
\makeatletter
\def\ps@pprintTitle{%
    \let\@oddhead\@empty
    \let\@evenhead\@empty
    \def\@oddfoot{\footnotesize\itshape
         {Article accepted to Computational Materials Science, \href{https://doi.org/10.1016/j.commatsci.2019.109418}{doi:10.1016/j.commatsci.2019.109418}} \hfill\today}%
    \let\@evenfoot\@oddfoot
    }
\makeatother

\usepackage[utf8]{inputenc}

\usepackage{hyperref}
\usepackage{natbib}
\usepackage{amsmath}
\usepackage{graphicx}
\graphicspath{{}}
\usepackage{cleveref}
\usepackage{lineno}
\usepackage{amssymb}
\usepackage{setspace}
\begin{document}
\title{Simulation of Coarsening in Two-phase Systems with Dissimilar Mobilities}

\author[1]{W. Beck Andrews}
\ead{wband@umich.edu}
\author[2]{Peter W. Voorhees}
\ead{p-voorhees@northwestern.edu}
\author[1]{Katsuyo Thornton\corref{cor1}}
\ead{kthorn@umich.edu}

\cortext[cor1]{Corresponding author}

\address[1]{Department of Materials Science and Engineering, University of Michigan, Ann Arbor, MI 48109, USA}
\address[2]{Department of Materials Science and Engineering, Northwestern University, Evanston, IL 60208, USA}

\begin{abstract}
In this work, we apply phase field simulations to examine the coarsening behavior of morphologically complex two-phase microstructures in which the phases have highly dissimilar mobilities, a condition approaching that found  in experimental solid-liquid systems.
Specifically, we consider a two-phase system at the critical composition ($50\%$ volume fraction) in which the mobilities of the two phases differ by a factor of 100.
This system is simulated in two and three dimensions using the Cahn-Hilliard model with a concentration-dependent mobility, and results are compared to simulations with a constant mobility.
A morphological transition occurs during coarsening of the two-dimensional system (corresponding to a thin film geometry) with dissimilar mobilities, resulting in a system of nearly-circular particles of high-mobility phase embedded in a low-mobility matrix.
This morphological transition causes the coarsening rate constant to decrease over time, which explains why a previous study found lack of agreement with the theoretical $t^{1/3}$ power law.
Three-dimensional systems with dissimilar mobilities resulted in bicontinuous microstructures that evolve self-similarly, as determined by quantitative analysis of the interfacial shape distribution.
Coarsening kinetics in three dimensions agreed closely with the $t^{1/3}$ power law after the initial transient stage.
A model is derived to explain a nearly-linear relationship between the coarsening rate constant and the variance of scaled mean curvature that is observed during this transient stage.

\end{abstract}


\maketitle
%\linenumbers
\section{Introduction}
Coarsening, or Ostwald ripening, is a fundamental physical process in which the microstructure of a system evolves to reduce the energy associated with its interfacial area.
The interfacial area can decrease when small microstructural features disappear, resulting in an increase in the overall length scale.
Coarsening is well-understood for systems of spherical particles at low volume fractions \cite{Baldan2002, Ratke2002}.
Theories originating with Lifshitz and Slyozov \cite{Lifshitz1961} and Wagner \cite{Wagner1961} (collectively known as LSW) predict morphologies and kinetics of self-similar evolution \cite{Brailsford1979, Marqusee1983, Yao1993}, where the microstructure becomes time independent when scaled by the characteristic length scale.
However, complex microstructures, including those observed during the coarsening of dendritic mixtures during casting \cite{Karma2001, Mendoza2003, Mendoza2004, Fife2009, Fife2014, Shahani2015}, are not well-described by those theories \cite{Marsh1996}.
Advances in computing power and in experimental and computational techniques have made it possible to directly examine the coarsening of complex microstructures in 3-D \cite{Mendoza2003,Fife2009}.
Yet the difficulty of modeling these structures analytically means that simulations remain our main source of theoretical insight into coarsening dynamics.
In particular, coarsening of complex microstructures has been examined using phase field simulations \cite{Kwon2007, Kwon2009, Kwon2010, Geslin2015}.
Such studies have found self-similarly evolving morphologies for highly idealized systems, with a constant mobility throughout the system and isotropic interfacial energies \cite{Kwon2007, Kwon2009, Kwon2010}.
However, most experimental system have two phases with dramatically different mobilities, such as in solid-liquid systems \cite{Mendoza2003, Mendoza2004, Fife2009, Fife2014, Shahani2015, Sun2017, Sun2018}.
To help bridge this gap, we examine how dissimilar mobilities of two phases affects the coarsening of an otherwise idealized system.

During coarsening, the microstructure evolves as a result of mass transport via diffusion driven by gradients in chemical potential (and corresponding gradients in concentration).
These gradients result from differences in curvature among nearby interfaces within the microstructure, since chemical potential is related to interfacial curvature by the Gibbs-Thomson effect \cite{Ratke2002, Herring1951book}.
The theoretical power law for scaling (obtained for self-similar evolution) during coarsening via bulk diffusion is $L \propto t^{1/3}$, where $L$ is a characteristic length scale of the evolving structure \cite{Herring1950, Mullins1986, Kohn2002}.
The power law obtained by LSW for self-similar coarsening in the low volume fraction limit \cite{Lifshitz1961, Wagner1961} is a special case of this result.
For this limit, LSW were also able to derive the coarsening rate and the particle size distribution corresponding to self-similar evolution.
These predictions allow direct comparison to experimental microstructures and determination of interfacial energies based on coarsening rate \cite{Ardell1966}.
Subsequent works (see reviews \cite{Baldan2002, Ratke2002}) have predicted self-similar particle size distributions at higher volume fractions.
On the other hand, complex geometries lack the spherical symmetry that enables analytical modeling in particulate systems.
Even when the evolution of a particular feature (such as a pair of dendrite arms \cite{Reeves1971} or a neck connecting regions of solid phase \cite{Aagesen2010,Aagesen2011}) can be modeled in a simplified way, the relationship between the evolution of specific features and that of the overall microstructure is not straightforward.
It is, however, possible to consider complex microstructures through statistical distributions of interfacial velocity and curvature on the interface.
Using this approach, DeHoff \cite{DeHoff1991} derived expressions for the evolution of global quantities, such as interfacial area and average mean curvature.
Statistical relationships between local interfacial velocity and curvature have also been examined empirically in experimental \cite{Fife2014} and simulated \cite{Park2017} systems.

While these advances have improved our understanding of coarsening in complex microstructures, they have not enabled prediction of self-similarly evolving complex morphologies.
Currently, the only approach that can yield predictions of self-similar morphologies is simulation, such as those based on the phase field method \cite{Kwon2007,Kwon2009,Kwon2010,Park2017,Park2015AC}.
In this paper, following previous work on coarsening with constant mobility \cite{Kwon2007,Kwon2009,Kwon2010,Park2017}, we examine the effect of phases with dissimilar mobilities.
We employ the Cahn-Hilliard model \cite{CahnHilliard1958, Cahn1961}, a diffuse interface model that has been widely used to simulate phase separation and coarsening \cite{Kwon2007,Kwon2009,Kwon2010,Park2017,Park2015AC, Rogers1988, Rogers1989, Lacasta1992, Sappelt1997,Sappelt1997EPL, Ahluwalia1999, Zhu1999, Garcke2003, Sheng2010, Ju2015, Dai2016, Zhang2016}.
Correspondence of the Cahn-Hilliard model to the sharp interface problem describing coarsening has been demonstrated via asymptotic analysis \cite{Pego1989}.

To numerically implement dissimilar mobilities of the two phases, we employ a concentration-dependent mobility.
Such an approach was originally used to study phase separation via surface or interfacial diffusion \cite{Lacasta1992, Zhu1999} and spinodal decomposition with a glassy phase \cite{Sappelt1997,Sappelt1997EPL,Ahluwalia1999}.
This approach has also been taken in previous two-dimensional (2-D) studies of coarsening with dissimilar mobilities \cite{Sheng2010,Ju2015,Dai2016}.
Sheng et al.\ \cite{Sheng2010} studied coarsening in both the one-sided (zero mobility in one phase) and dissimilar-mobility cases.
They found a fitted coarsening exponent of $1/3.3$  (i.e., $L\propto t^{1/3.3}$) and scaling of the pair correlation function, a measure of morphology, by the first moment of the structure function, one of the characteristic length scales they examined.
Similarly, Ju et al.\ \cite{Ju2015} reported a fitted coarsening exponent of $1/3.2$ for the one-sided case.
However, in contrast to those results, Dai and Du predicted \cite{Dai2016} reported agreement with the $t^{1/3}$ power law.
In three dimensions, simulations of coarsening with a one-sided mobility were recently carried out \cite{Zhang2016}, but the resulting morphologies were not characterized.

In this work, we aim to understand how dissimilar mobilities affects the coarsening morphology and kinetics of complex microstructures in two and three dimensions (2D and 3D, respectively).
We present a general phase field model for evolution via bulk diffusion of a two-phase system with dissimilar mobilities.
Then, we present simulations with dissimilar mobilities in 2D and 3D with large sample microstructures.
Coarsening kinetics and morphological evolution are analyzed, and morphologies are quantified using statistical measures such as the interfacial shape distribution (ISD) \cite{Mendoza2004}.
Results with constant mobility are presented for comparison.
In the 2-D case, our results address the question of whether coarsening dynamics follow the theoretical $t^{1/3}$ power law.
In the three-dimensional (3-D) case, we relate morphology and kinetics by applying statistical assumptions to the relationship between interfacial velocity and local mean curvature.

\section{Model \label{sec:model}}
In order to study coarsening in systems where the phases have dissimilar mobilities, we employ the Cahn-Hilliard equation with concentration-dependent mobility.
The thermodynamic basis for this model is introduced in Refs.\ \cite{CahnHilliard1958} and \cite{Cahn1961}.
The Cahn-Hilliard equation is
\begin{equation}
\label{eq:CH}
    \frac{\partial \phi}{\partial t} = \nabla \cdot M(\phi) \nabla \mu,
\end{equation}
where $\phi$ is the scaled concentration and $M(\phi)$ is the mobility. (See Appendix A for details regarding scaling/rescaling of $\phi$).
The chemical potential $\mu$ is
\begin{equation}
\label{eq:CH_mu}
    \mu = f'(\phi) -\epsilon^2 \nabla^2 \phi,
\end{equation}
where $\epsilon$ is the gradient energy coefficient and $f(\phi)$, the bulk free energy, is given by
\begin{equation}
\label{eq:bulken}
    f(\phi) = \frac{W}{4} \phi^2 (\phi - 1)^2.
\end{equation}
The minima of $f(\phi)$, $\phi_0^+=1$ and $\phi_0^- = 0$, correspond to the equilibrium concentrations of two phases in contact at a planar interface.
The parameter $W$ controls the height of the energy barrier between the two energy minima.

To connect this model to the sharp interface dynamics employed in classical theories of coarsening, we note that the capillary length of the phase field model can be expressed as \cite{Ratke2002, Pego1989}
\begin{equation}
\label{eq:capillary}
    \Gamma = \frac{2\gamma}{f''(\phi_0) (\phi_0^- - \phi_0^+)^2},
\end{equation}
where the interfacial energy $\gamma$ is given by \cite{CahnHilliard1958}
\begin{equation}
\label{eq:gamma}
    \gamma = \epsilon \int_{\phi_0^-}^{\phi_0^+}\sqrt{2f(\phi)}d\phi.
\end{equation}
The Gibbs-Thomson equation for the equilibrium chemical potential at a curved interface is then
\begin{equation}
\label{eq:GT_mu}
    \mu = \frac{2 \gamma H}{\phi |^+_-},
\end{equation}
where $H$ is mean curvature and $\phi |^+_-$ indicates the difference in concentration between the phases.
The resulting equilibrium condition for $\phi$ is
\begin{equation}
\label{eq:GT_phi}
    \phi^\pm = \phi_0^\pm + \Gamma H,
\end{equation}
where $H$ is mean curvature.
Motion of the interface is specified by a mass conservation condition,
\begin{equation}
\label{eq:interfacial_velocity}
    v = -\frac{ \vec n \cdot \vec j |_-^+ }{\phi |^+_-},
\end{equation}
where $v$ is the normal velocity of the interface, $\vec n$ is the interface normal vector, and $\vec j$ is the diffusive flux, $\vec j = - M(\phi) \nabla \mu$.
Chemical potential in the bulk of each phase is assumed to satisfy the Laplace equation
\begin{equation}
\label{eq:laplace}
    \nabla^2 \mu = 0,
\end{equation}
which describes steady-state diffusion between interfaces.

The form of the concentration-dependent mobility $M(\phi)$ determines its sensitivity to changes in concentration.
Following the approach taken in phase field modeling of solid-liquid systems \cite{Karma2001, Warren1995}, we express $M(\phi)$ in the general form
\begin{equation}
\label{eq:mobility}
    M(\phi) = (M^+ - M^-) h(\phi) + M^-,
\end{equation}
where $M(\phi_0^+)=M^+$ and $M(\phi_0^-)=M^-$ are the desired bulk mobilities ($M^+>M^-$ in this study), and $h(\phi)$ is a smooth interpolation function that satisfies $h(\phi_0^+)=1$ and $h(\phi_0^-)=0$.
Changes in mobility resulting from deviation in concentration in the bulk (e.g., due to the Gibbs-Thomson effect) can affect dynamics if the corresponding bulk mobility is set to be small \cite{BrayEmmott1995,DaiDu2012, Lee2016siam}.
Bray and Emmott \cite{BrayEmmott1995} describe how this effect can modify the classical LSW coarsening dynamics of a system of particles, while Lee et al.\ \cite{Lee2016siam} consider it numerically and analytically for a single particle.
Refs.\ \cite{BrayEmmott1995} and \cite{Lee2016siam} consider models where bulk mobility is set to zero in both phases (i.e., models for surface diffusion), but Dai and Du \cite{DaiDu2012} show that this effect can also occur with dissimilar mobilities.
In our model, the form of $h(\phi)$ determines how much $M(\phi)$ changes when $\phi$ deviates from its equilibrium values, $\phi_0^\pm$.
We therefore select the polynomial interpolation function,
\begin{equation}
\label{eq:interp}
    h(\phi) = \phi^3 (10-15\phi+6\phi^2),
\end{equation}
to sharpen the transition of the mobility from $M^-$ to $M^+$ near $\phi=0.5$.
As in the sinusoidal interpolation used in Ref.\ \cite{Sheng2010}, this polynomial form reduces the sensitivity of $M(\phi)$ to the deviation of $\phi$ from its bulk values compared to linear interpolation, which has been used more commonly \cite{Sheng2010,Ju2015,Dai2016}.

\begin{figure}
	\centering
	\includegraphics[width = 7.8cm]{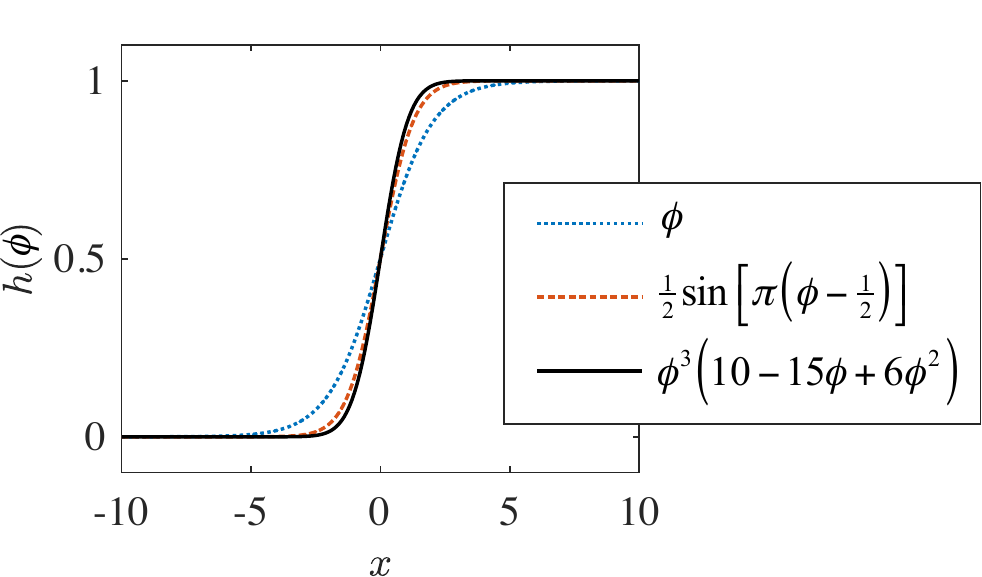}
	\caption{
		Comparison of interpolation functions $h(\phi)$ using the analytical interfacial profile $\phi(x) = \frac{1}{2}\left[1+\tanh(x/2) \right]$ (blue dotted curve).  The black solid curve is the interpolation function employed in this work. The interpolation function using a sine function from Ref.\ \cite{Sheng2010} is also shown by the red dashed curve.
	}
	\label{fig:interp}
\end{figure}

The effect of the choice of interpolation function on the mobility of the low-mobility phase can be demonstrated by considering a small difference in concentration $\omega$, where $\phi = \phi_0^- + \omega$, and expanding the mobility in a Taylor series about $\phi_0^-$.
For linear $h(\phi)$, $M(\phi) = M^- - \omega h'(\phi_0^-) + O(\omega^2)$.
For our choice of $h(\phi)$, $M(\phi) = M^- - \frac{1}{6}\omega^3 h'''(\phi_0^-) + O(\omega^4)$.
That is, our choice of interpolation function reduces the difference between the desired and actual bulk mobilities, $|M(\phi)-M^-|$, to $O(\omega^3)$ because it satisfies $h'(\phi_0^\pm) = h''(\phi_0^\pm)=0$.
In principle, $|M(\phi)-M^-|$ could be reduced to arbitrary order $m$ by requiring that $h^{(m)}(\phi_0^\pm)=0$, but increasing $m$ results in a sharper transition of $h(\phi)$ through the interface (see Fig.\ \ref{fig:interp}).
The interpolation function in Eq.\ \ref{eq:interp} represents a compromise between being able to smoothly resolve $M(\phi)$ through the interface and reducing $|M(\phi)-M^\pm|$ in the bulk, which is important when one of the mobilities is small.

\section{Numerical and Characterization Methods}
Our simulations are conducted by evolving $\phi$ according to Eqs.\ \ref{eq:CH} and \ref{eq:CH_mu} with periodic boundary conditions in two and three spatial dimensions, using the bulk free energy given by Eq.\ \ref{eq:bulken} and the concentration-dependent mobility defined by Eqs.\ \ref{eq:mobility} and \ref{eq:interp}.
We use parameters $\epsilon^2 = 0.2$ and $W=0.4$, which yield $\gamma = 1/30$ per Eq.\ \ref{eq:gamma}.
We set $M^+=1$ and $M^-=10^{-2}$ for the simulations with dissimilar mobilities, and $M=1$ for the simulations with constant mobility.
All of these parameters are dimensionless, and the procedure for nondimensionalization is described in Appendix A.

Finite differences were employed to discretize Eqs.\ \ref{eq:CH} and \ref{eq:CH_mu} in space and time.
The time discretization consisted of an explicit forward Euler scheme with $\Delta t = 0.05$, which is near the stability limit in 3D.
The computational domain is discretized into a uniform grid with $\Delta x = 1$, which results in 3-5 grid points in the interface.
Centered differences were used to approximate spatial derivatives, including the conventional five-point (2D)/seven-point (3D) stencil to approximate the Laplacian.
The term $\nabla \cdot M(\phi)\nabla \mu$ in Eq.\ \ref{eq:CH} was approximated by computing $M(\phi)\nabla \mu$ on half-points (i.e., halfway between grid points), and taking a centered difference to find the divergence $\nabla \cdot M(\phi)\nabla \mu$ on the grid points.
The mobility $M(\phi)$ was approximated on half-points by averaging $M(\phi)$ calculated at the grid points, and the first derivatives composing $\nabla \mu$ were approximated on half-points by centered differences.
The finite difference scheme was implemented in Fortran 90, and the simulation code was parallelized with MPI.
A 3-D dissimilar-mobilities simulation with $1024^3$ grid points took 25 hours for $10^6$ iterations (1/8th of the total simulated duration) on 256 cores of the TACC Stampede supercomputer.
2-D simulations were conducted with a slightly modified version of the 3-D code, and the 2-D dissimilar-mobilities simulation (with $2844^2$ grid points) took 37 hours for $1.28\times10^7$ iterations on 64 Intel Haswell cores of the Flux computing cluster at the University of Michigan.

Simulated structures were characterized quantitatively in terms of characteristic length and the interfacial shape distribution.
We employ the inverse of the interfacial area per unit volume,  $S_V^{-1}$, to define the characteristic length of the simulated structures \cite{Marsh1996}.
We also use $S_V^{-1}$ to denote the analogous characteristic length in two dimensions, the ratio of area to interfacial length.
The interfacial shape distribution (ISD) of the interface of a 3-D structure is the probability density function for the probability that a point  on its interface has a specific pair of principal curvatures, $\kappa_1$ and $\kappa_2$ \cite{Mendoza2003}.
We express it as
\begin{equation}
    \label{eq:ISD3D}
    P(\kappa_1,\kappa_2) = \frac{1}{A_T} \frac{\partial^2 F_A (\kappa_1,\kappa_2)}{\partial \kappa_1 \partial \kappa_2},
\end{equation}
where $A_T$ denotes the total interfacial area of the structure and $F_A(\kappa_1, \kappa_2)$ is a cumulative area distribution function, defined as the total area within the structure having first and second principal curvatures less than or equal to $\kappa_1$ and $\kappa_2$, respectively.
Equation \ref{eq:ISD3D} is equivalent to definitions of the ISD in Refs.\ \cite{Mendoza2003} and \cite{Park2017}, but the expression based on the differential of the cumulative area distribution provided here is advantageous for the derivations in Appendix C.
Analogously to the 3-D case, we define the ISD of a 2-D structure as the probability density function for the probability of a point on the interface having a specific scalar curvature $\kappa$, i.e.,
\begin{equation}
    P(\kappa) = \frac{1}{s_T} \frac{d F_s (\kappa)}{d \kappa},
\end{equation}
where $s_T$ is the total interfacial length and $F_s$ is the cumulative interfacial length within the structure having scalar curvature less than or equal to $\kappa$.

During self-similar evolution, an ISD calculated with curvatures scaled by the evolving characteristic length, e.g., $\kappa_1/S_V$,  must be time-invariant.
The scaled ISD is therefore a useful tool to evaluate whether evolution is in fact self-similar and to observe changes in morphology independent of the change in length scale.
All ISDs presented in this paper are scaled by $S_V^{-1}$.
Processing of the simulation data and calculation of interfacial curvatures were conducted as described in Ref.\ \cite{Park2014}.
The sign convention for mean curvature is such that low-mobility-phase convex bodies (e.g., spheres) have positive curvatures.

In 2D, we perform one simulation with constant mobility and one simulation with dissimilar mobilities.
In 3D, we perform one simulation with constant mobility and two simulations with dissimilar mobilities, one with dissimilar mobilities for the entire simulation, and another with constant mobility for phase separation and dissimilar mobilities for subsequent coarsening.
We chose to conduct a single simulation for each setup (rather than multiple simulations with smaller domains) to ensure that the simulation domains contained statistically representative structures and that the results are not affected by the periodic boundary conditions imposed on the domain boundaries over the longer simulation times required to examine coarsening. 
Our simulations were initialized with random noise, and therefore increasing domain size should have the same effect on statistical significance as the averaging of multiple simulations, and the larger domain has the advantage of reduced boundary effects.

\section{2-D Simulations}
Two 2-D simulations were conducted, one with constant mobility and one with dissimilar mobilities, in square domains with sides of length $L_x = L_y = 2844$.
Both were initialized with the same random concentration values uniformly distributed within the interval $0.40-0.60$.
This results in volume fractions of $50\%$ for both phases at late times, although in the dissimilar-mobilities structure more low-mobility phase is initially present due to its slower rejection of solute.

\subsection{Morphology}
\begin{figure}
	\centering
	\includegraphics[width = 14.4cm]{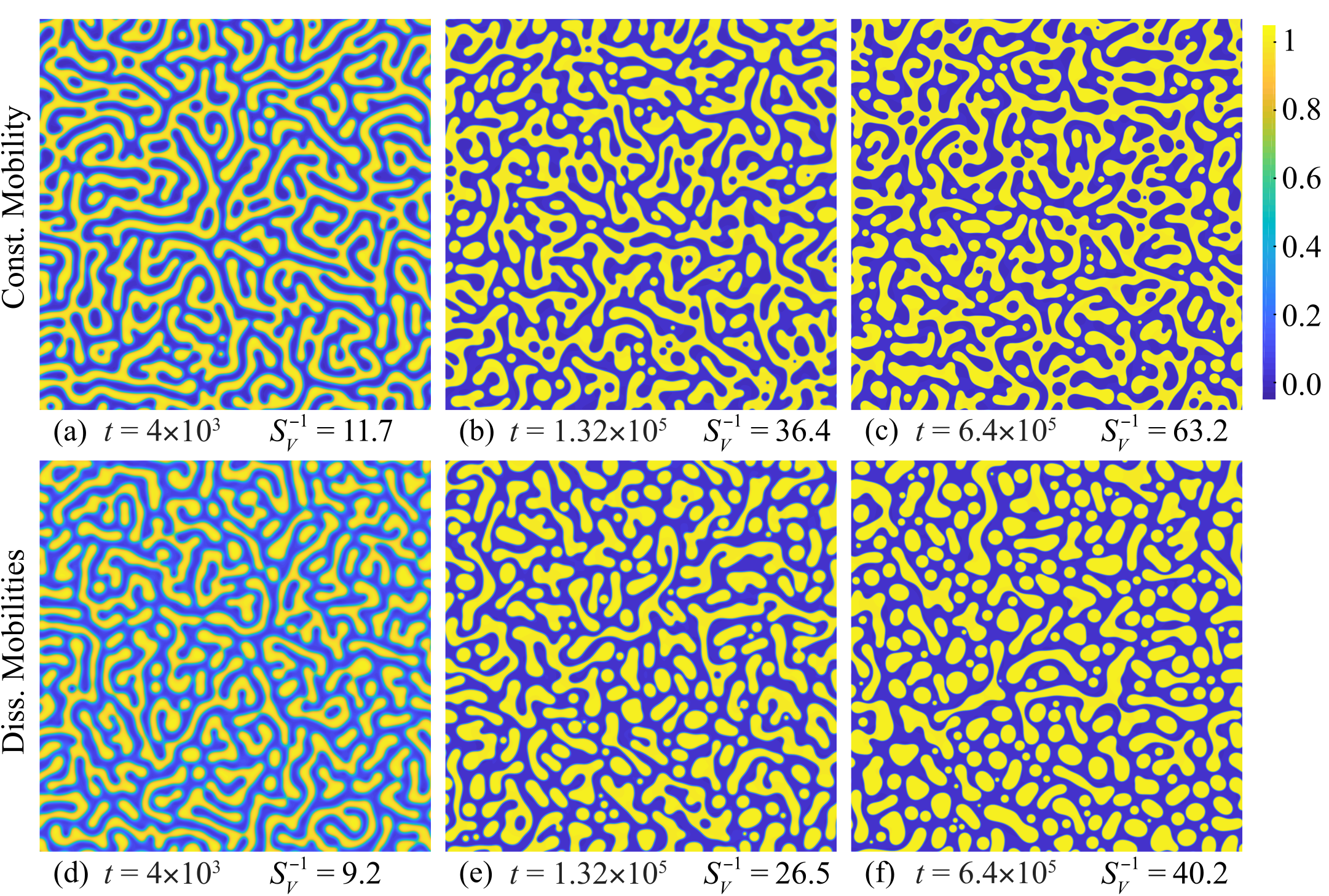}
	\caption{
		Evolution of morphologies in 2D during coarsening with constant mobility (a-c) and dissimilar mobilities (d-f).  Each subfigure depicts $\phi$ within a square subdomain with side length $45S_V^{-1}$.  In (d-f), blue indicates the low-mobility phase, while yellow corresponds to the high-mobility phase.
	}
	\label{fig:2D_struct}
\end{figure}

Figure \ref{fig:2D_struct} depicts the morphology of the simulated 2-D structures during coarsening.
The constant mobility structure is shown in Figs.\ \ref{fig:2D_struct}a-c, and the dissimilar mobilities structure is shown in Figs.\ \ref{fig:2D_struct}d-e.
The portions of the simulated domain presented in Fig.\ \ref{fig:2D_struct} are chosen to have the same characteristic area (with the side length $45S_V^{-1}$), enabling us to directly observe morphological changes without the effects of the change in $S_V^{-1}$.
Figures \ref{fig:2D_struct}a and \ref{fig:2D_struct}d show the structures at $t=4\times10^3$, near the end of phase separation.
Both structures appear to consist of alternating layers of each phase.
While in many locations the layers appear to be flat, they bend and terminate such that the isotropy of the overall structures is preserved.
The dissimilar-mobility structure appears to have more high-mobility-phase particles than low-mobility-phase particles, while in the constant-mobility structure, the phases appear to have statistically the same morphology.

As the simulations progress, the constant-mobility structure forms large percolating regions of each phase that contain smaller regions of the opposite phase (Figs.\ \ref{fig:2D_struct}b and \ref{fig:2D_struct}c).
The initially layered structures that were dominated by nearly flat interfaces evolve to more sinuous ones.
Overall, these morphological changes are subtle.
In contrast, the dissimilar-mobility structure undergoes a significant morphological transition from the initial layered structure to a structure consisting of high-mobility-phase particles in a low-mobility matrix (Figs.\ \ref{fig:2D_struct}e and \ref{fig:2D_struct}f).
This transition occurs through both the disappearance of low-mobility-phase particles and the evolution of high-mobility-phase regions toward their circular equilibrium shape.
High-mobility-phase particles also disappear and coalesce, but not rapidly enough to reduce their predominance in the overall structure.

\begin{figure}
	\centering
	\includegraphics[width = 12.3cm]{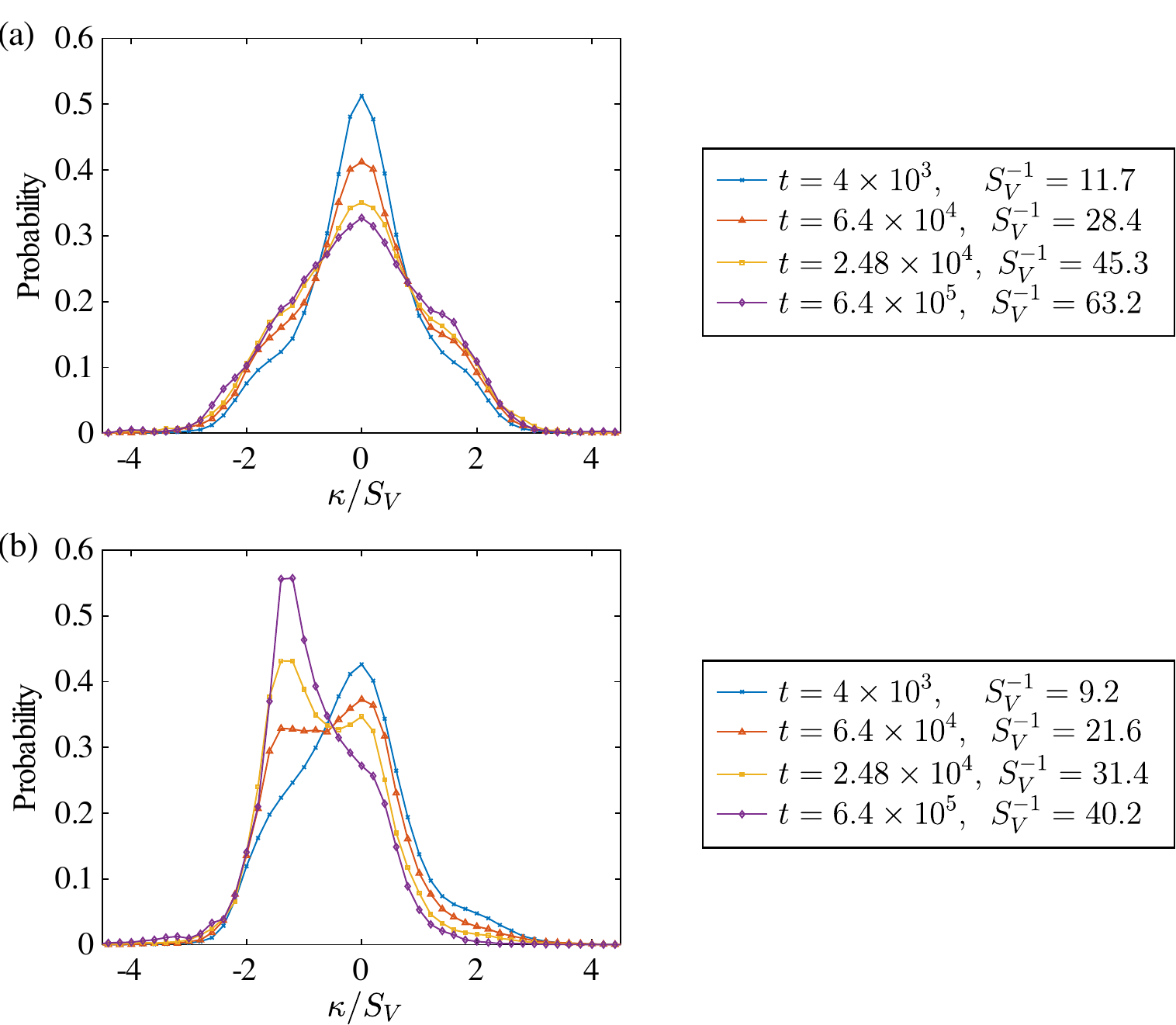}
	\caption{
		Time evolution of interfacial shape distributions (ISDs) of the 2-D structures, (a) coarsened with constant mobility, and (b) coarsened with dissimilar mobilities.  ISDs are shown corresponding to four times, $t=4\times 10^3$ (blue curve/x-symbols), $t=6.4\times 10^4$ (red curve/triangles), $t=2.48\times 10^5$ (yellow curve/squares), and $t=6.4\times 10^5$ (purple curve/diamonds).
	}
	\label{fig:2D_ISD}
\end{figure}

The morphological evolution seen in the simulated structures is quantified by the changes in the ISDs, which are shown in Fig.\ \ref{fig:2D_ISD}.
The nearly flat interfaces that are predominant in the layered morphology observed at $t=4\times 10^3$ correspond to peaks at $\kappa/S_V=0$ on the ISDs in Fig.\ \ref{fig:2D_ISD}.
In the constant-mobility case, the ISD (Fig.\ \ref{fig:2D_ISD}a) is symmetric about $\kappa/S_V=0$, which is consistent with the symmetry between the phases: they have the same volume fraction ($50\%$) and, with constant mobility, the same transport kinetics.
As the constant-mobility structure evolves, the peak of the ISD remains centered at $\kappa/S_V=0$, but it broadens over time, with standard deviation $\sigma_{\kappa/S_V}$ increasing from $1.0$ to $1.3$.
These changes can be understood in terms of the evolution of three overlapping populations in the ISD: one corresponding to the population having nearly flat interfaces centered at $\kappa/S_V=0$, and two symmetric populations that correspond to high-curvature features (particles and end-caps of layers) and appear between $\kappa/S_V=-2$ and $\kappa/S_V=-1$ and $\kappa/S_V=1$ and $\kappa/S_V=2$.
In the constant-mobility ISD, the populations of high-curvature features increase symmetrically, while the population with nearly flat interfaces decreases.
This may appear unintuitive based on the Gibbs-Thomson condition, but it is not a contradiction because the curvatures are scaled by the characteristic length scale, and overall the interfacial energy is lowered by the reduction of interfaces with large unscaled curvatures as well as the nearly flat interfaces.

The dissimilar-mobility ISD (Fig.\ \ref{fig:2D_ISD}b) is asymmetric at $t=4\times10^3$, with more interface having negative curvature than positive curvature.
Negative curvature corresponds to convex high-mobility-phase features, and the asymmetry of the $t=4\times10^3$ ISD is consistent with the greater prevalence of high-mobility-phase particles in Fig.\ \ref{fig:2D_struct}d.
The morphological transition observed in Figs.\ \ref{fig:2D_struct}d-f is represented in the ISD by the growth of the population centered around $\kappa/S_V = -1.3$ and by the decay of population with positive curvature.
The growth near $\kappa/S_V = -1.3$ is due to the longevity of high-mobility-phase domains surrounded by low-mobility-phase matrix observed in Figs.\ \ref{fig:2D_struct}e and \ref{fig:2D_struct}f.
The disappearance of the population with positive curvature corresponds to the loss of low-mobility-phase particles and the evolution of domains of high-mobility phase toward their equilibrium shape, as discussed earlier.
Based on the ISDs in Fig.\ \ref{fig:2D_ISD}b, evolution during this simulation is clearly not self-similar, and unlike the constant mobility case, the structure continues to undergo significant evolution even at the latest time we examined.
Thus, the 2-D dissimilar-mobilities structure remains within a transient regime for the entire duration of the simulation.

\subsection{Kinetics}
\begin{figure}
	\centering
	\includegraphics[width = 10.8cm]{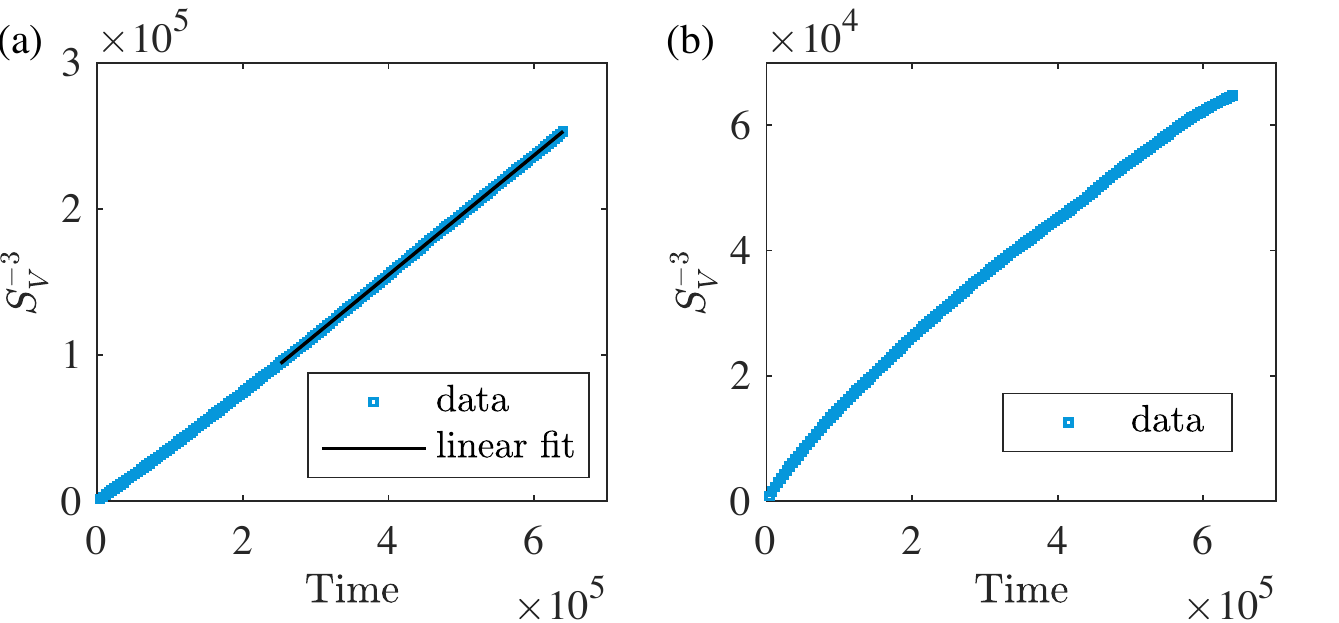}
	\caption{
		 Coarsening kinetics of the 2-D structures with (a) constant mobility and (b) dissimilar mobilities.  $S_V^{-3}$ is plotted vs.\ time (blue squares) to evaluate adherence to the power law, and a linear fit (solid black line) is provided for the constant-mobility case.
	}
	\label{fig:2D_kinetics}
\end{figure}

The theoretical $t^{1/3}$  power law for the characteristic length $S_V^{-1}$ is expressed as
\begin{equation}
\label{eq:powerlaw}
    S_V^{-3}(t) - S_V^{-3}(0) = kt,
\end{equation}
 where $k$ is the coarsening rate constant.
$S_V^{-3}$ is plotted vs.\ time in Figs.\ \ref{fig:2D_kinetics}a and \ref{fig:2D_kinetics}b for the constant-mobility and dissimilar-mobility cases, respectively.
In the constant-mobility case, the coarsening rate constant increases during an initial transient stage, consistent with the results of \cite{Garcke2003}.
At later times ($t > 2.48\times 10^5$), a linear fit was used to evaluate convergence to the power law in Eq.\ \ref{eq:powerlaw}.
Good agreement was found, with $R^2 = 0.99996$, and the equation of fit was $S_V^{-3} = 0.409t -8.93\times 10^3$.
In the dissimilar-mobility case, the coarsening rate constant $k$ appears to be decreasing with time over the course of the simulation.
Evaluating the coarsening rate constant at the beginning and end of the simulation ($t < 2\times 10^4$ and $6.2\times 10^5 < t$, respectively), we find that $k$ decreases from $0.181$ to $0.067$, a factor of $2.7$.
The coarsening rate will likely continue to decrease in this case until the scaled morphology reaches a steady state, which may be a particulate structure with circular high-mobility-phase domains embedded in the low-mobility phase.

\subsection{Discussion \label{sec:2D_diss}}
In the dissimilar-mobility case, the complex layered structure resulting from phase separation was observed to transform over time into a system of high-mobility-phase particles in a low-mobility matrix.
This transition can be explained in terms of the diffusive interactions between neighboring patches of interface.
Interfaces can interact through both phases, but an interface will only coarsen if it has different curvature than the interfaces it is interacting with.
We classify two types of geometric features: those that can coarsen via interactions through the high-mobility phase and those that cannot.
The first type consists of low-mobility-phase particles (which are surrounded by high-mobility phase) and non-circular high-mobility-phase regions, especially those with complex or elongated shapes.
The second type consists of circular high-mobility-phase particles surrounded by low-mobility phase, which can only evolve via interactions through the low-mobility phase because all of the interfaces exposed to the high-mobility phase have the same curvature.
The asymmetry in kinetics between these two types of features drives the morphologies observed in Fig.\ \ref{fig:2D_struct} and \ref{fig:2D_ISD}: low-mobility-phase particles disappear, complex high-mobility-phase particles become circular, and circular high-mobility-phase particles persist.

\begin{figure}
	\centering
	\includegraphics[width = 8.4cm]{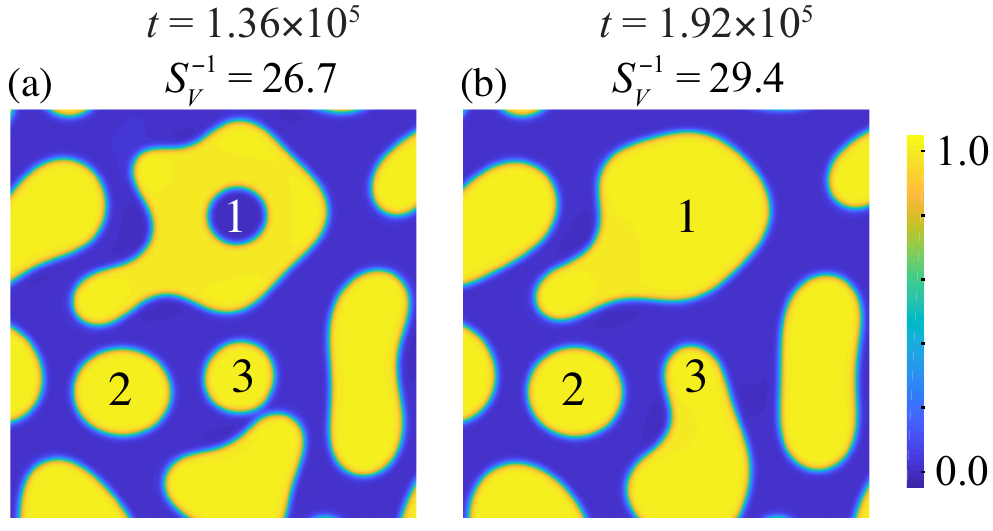}
	\caption{
	    Illustration of the evolution of different types of feature within the 2-D dissimilar-mobility structure.  The concentration field is shown within a small ($219^2$) part of the structure at times (a) $t=1.36\times 10^5$ and (b) $t=1.92\times 10^5$.  Three features are labeled: 1) a particle surrounded by high-mobility phase that completely disappears, 2) a particle surrounded by low-mobility phase that does not evolve, and 3) a particle surrounded by low-mobility phase that merges with a neighboring high-mobility-phase particle.
	}
	\label{fig:2D_example}
\end{figure}

The evolution of specific features is illustrated in Fig.\ \ref{fig:2D_example}, which shows a small region of the dissimilar-mobilities structure before and after a small amount of evolution ($S_V^{-1}$ changes from 26.7 to 29.4 between Figs.\ \ref{fig:2D_example}a and \ref{fig:2D_example}b).
Feature 1 in Fig.\ \ref{fig:2D_example}a is a low-mobility particle surrounded by high-mobility phase (i.e., a feature of the first type described above).
This particle evolves very rapidly, disappearing completely between Figs.\ \ref{fig:2D_example}a and \ref{fig:2D_example}b.
Feature 2 is a nearly-circular high-mobility-phase particle (i.e., a feature of the second type), and it is essentially unchanged between Figs.\ \ref{fig:2D_example}a and \ref{fig:2D_example}b.
Feature 3 is also a circular high-mobility-phase particle, but it coalesces with a nearby particle into a more complex shape that evolves rapidly.
These coalescence events create new features of the first type (complex or elongated high-mobility particles) out of other high-mobility-phase particles of either type.
However, coalescence events are not frequent enough (at the volume fraction considered here) to prevent circular high-mobility-phase particles from becoming the most prevalent feature in the structure by the end of the simulation.

The morphological transition described above explains the observed decrease in coarsening rate constant.
The area fraction of the circular high-mobility-phase particles surrounded by low-mobility phase increases over time because their evolution is controlled by the diffusion across the low-mobility phase, leading to slower evolution than that controlled by the diffusion across the high-mobility phase. 
Sheng et al.\ \cite{Sheng2010} found $L\propto t^{1/3.3}$ instead of the theoretically predicted $t^{1/3}$ power law, which is consistent with a decrease in the coarsening rate constant ($k$ in Eq.\ \ref{eq:powerlaw}) over time.
We observe a decrease in $k$, although we do not attribute it to a change in the underlying power law, which is based on the scaling of the governing equations.
In contrast, Dai and Du \cite{Dai2016} reported agreement with the $t^{1/3}$ power law.
Comparing their results to ours with the appropriate rescaling (detailed in Appendix A), we find that they fit to the coarsening kinetics within a timescale equivalent to $t\le 6.4 \times 10^4$ with our parameters.
Agreement with the $t^{1/3}$ power law within this early timescale is consistent with our results since the decrease in coarsening rate constant is more evident at later times.

The morphological transition would have been difficult to observe without our use of the ISD to characterize the morphologies of our simulated structures.
The structure function and pair correlation function, employed by Sheng et al.\ \cite{Sheng2010} to characterize their structures, were previously found to be insensitive to the difference between complex percolating domains and particles embedded in a matrix \cite{Rogers1989}.
Those types of morphology are clearly differentiated in the ISD, and we were able to observe a transition from one type to the other.
Thus, we conclude that the classical 1/3 exponent is not observed in the dissimilar-mobility simulation because the two-phase mixture is not yet self-similar, i.e., it is still within a transient regime.

\section{3-D Simulations}
Three 3-D simulations were conducted with domains of size $L_x = L_y = L_z = 1024$, initialized  with $\phi$ uniformly distributed within $0.40-0.60$ and generated with the same set of random numbers.
The first simulation used constant mobility at all times.
The second simulation, dissimilar mobilities PS IC (phase-separated initial condition), employed constant mobility until $t=10^4$ when phase separation was complete, and dissimilar mobilities for subsequent coarsening.
The third simulation, dissimilar mobilities RN IC (random noise initial condition), was conducted with dissimilar mobilities for the duration of the simulation, $0 < t \le 4\times 10^5$.
This resulted in a higher initial volume fraction of low-mobility phase (similarly to the 2-D dissimilar-mobility case), which converged over time to 50\%.
These two simulations (PS IC and RN IC) were conducted for the dissimilar-mobility case in 3D in order to determine whether the self-similar state is initial-condition dependent, which was not necessary in the 2-D case because of the lack of a self-similar state.

\subsection{Morphology}
\begin{figure}
	\centering
	\includegraphics[width = 10.8cm]{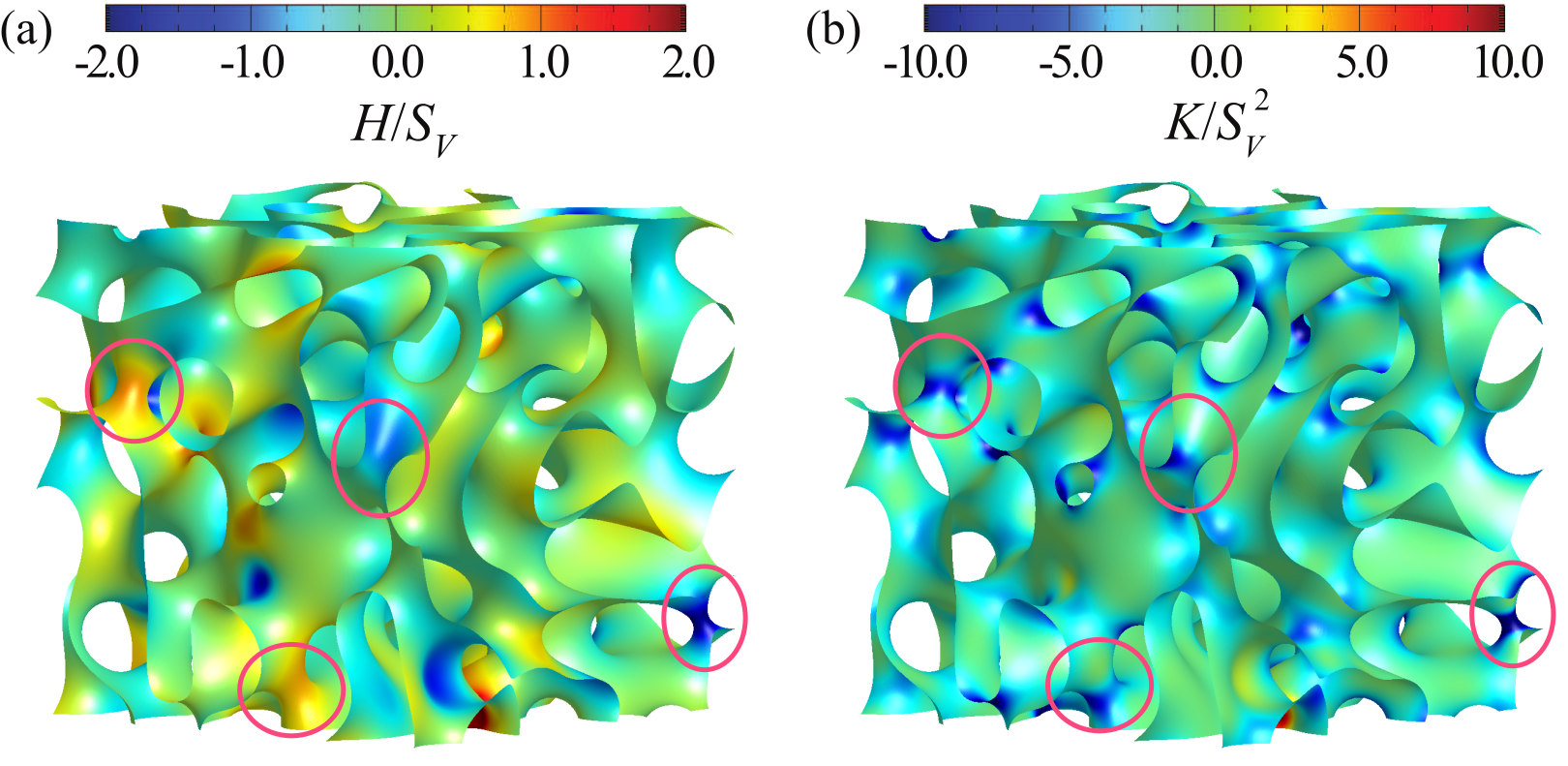}
	\caption{
		Late-time interfacial morphology for the dissimilar-mobility PS IC case at time $t=4\times 10^5$.  The $\phi=0.50$ iso-surface is shown colored by (a) scaled mean curvature and (b) scaled Gaussian curvature within a cubic subdomain with side length $8S_V^{-1} = 283$. Four necks are circled, two of which contain high-mobility phase (negative $H/S_V$) and two of which contain low-mobility phase (positive $H/S_V$).
	}
	\label{fig:3D_struct}
\end{figure}

All three simulation conditions resulted in qualitatively similar bicontinuous microstructures.
A representative morphology is shown in Fig.\ \ref{fig:3D_struct} for the dissimilar-mobility case.
Specifically, Fig.\ \ref{fig:3D_struct} shows the $\phi=0.50$ isosurface of the dissimilar-mobility PS IC structure at time $t=4\times 10^5$ within a cubic subdomain with side length $8S_V^{-1}$.
The isosurface is colored by scaled mean and Gaussian curvature in Figs.\ \ref{fig:3D_struct}a and \ref{fig:3D_struct}b, respectively.
The predominant high-curvature features are necks, four of which are circled.
These necks have large negative Gaussian curvatures.
Mean curvatures can be positive or negative depending on which phase the neck contains: positive for necks containing low-mobility phase and negative for necks containing high-mobility phase.
Negative Gaussian curvature indicates that interfaces are hyperbolic, i.e., they have oppositely signed principal curvatures $\kappa_1$ and $\kappa_2$.
As evident in Fig.\ \ref{fig:3D_struct}, most of the interfaces are either hyperbolic or nearly planar.
Elliptic interfaces (with principle curvatures of the same sign) with large positive Gaussian curvatures are also present.
However, they are rare and are most likely products of the pinching of necks, which disappear relatively quickly because the same-signed principle curvatures add to the magnitude of the mean curvature that determines the driving force for evolution via the Gibbs-Thomson condition.

\begin{figure}
	\centering
	\includegraphics[width = 14.4cm]{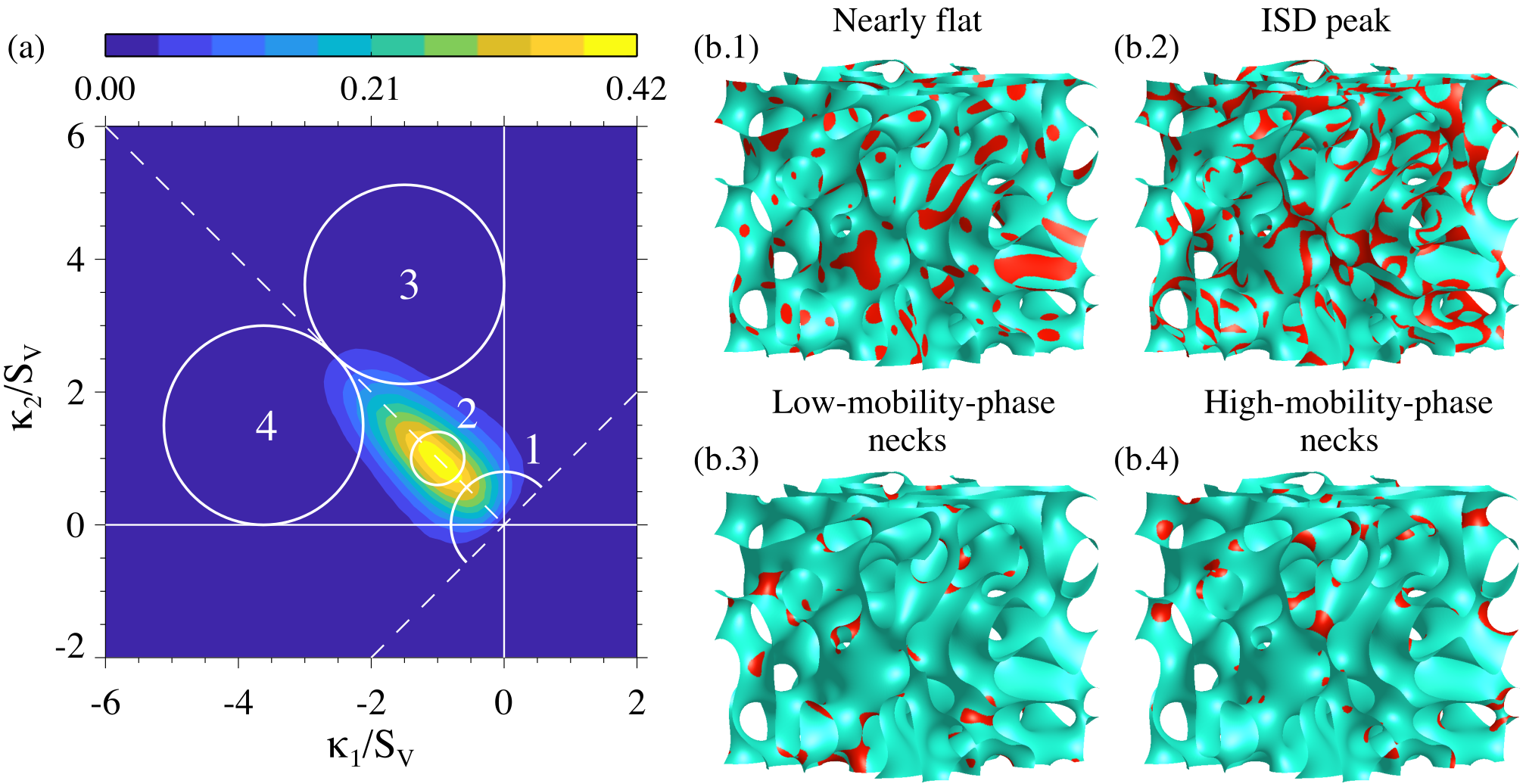}
	\caption{
	  (a) ISD and (b) structures identifying four different types of interface: 1) nearly flat interfaces, 2) interfaces at the ISD peak, 3) high-mobility-phase necks, and 4) low-mobility-phase necks.  The ISD and structures shown are for the dissimilar-mobility PS IC case at time $t=4\times 10^5$.
	}
	\label{fig:3D_ISDtypes}
\end{figure}

To consider how populations of different types of interface compare between structures and evolve in time, we define four types of interfaces based on regions of the ISD shown in Fig.\ \ref{fig:3D_ISDtypes}a.
The first two types of interface are considered due to their location within the ISD: 1) nearly flat interfaces, in which $|\kappa_1|$ and $|\kappa_2|$ are both near zero, and 2) interfaces near the ISD peak, $(\kappa_1,\kappa_2)\approx (-1,1)$.
The third and fourth types are intended to correspond to the necks identified in Fig.\ \ref{fig:3D_struct}: 3) necks containing low-mobility phase and 4) necks containing high-mobility phase.
Thus, 3) and 4) both have large positive $\kappa_2$ and large negative $\kappa_1$, but 3) has positive mean curvature (i.e., $\kappa_2>-\kappa_1$), while 4) has negative mean curvature.
In Figs.\ \ref{fig:3D_ISDtypes}b.1-4, the four types of interface are illustrated by red highlights  on the dissimilar-mobilities structure from Fig.\ \ref{fig:3D_struct}.
In Fig.\ \ref{fig:3D_ISDtypes}b.1, we see that nearly flat interfaces are present on the structure in small, round patches and larger non-circular areas.
Interfaces near the ISD peak, highlighted in Fig.\ \ref{fig:3D_ISDtypes}b.2, are present in thin strips of area, some of which appear to partially enclose areas of nearly flat interfaces.
As expected, Figs.\ \ref{fig:3D_ISDtypes}b.3 and b.4 show that regions 3) and 4) of the ISD correspond primarily to necks in the structures, although some neck areas in Fig.\ \ref{fig:3D_struct} are out of the limits of regions 3) and 4) and thus are not highlighted.

\begin{figure}
	\centering
	\includegraphics[width =13.6cm]{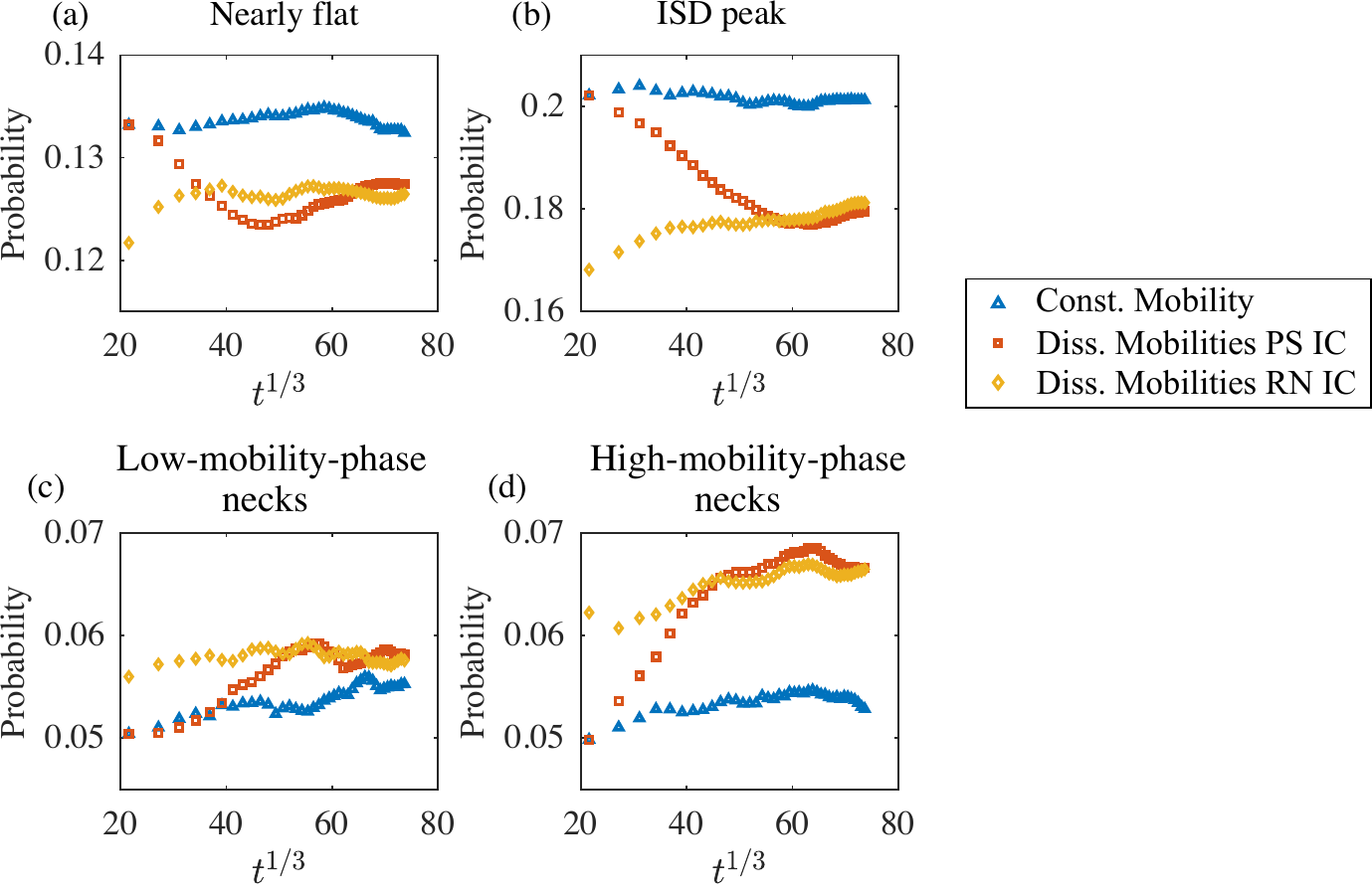}
	\caption{
		Time evolution of the integrated probability (i.e., the area fraction) of the four types of interfacial shape identified in Fig.\ \ref{fig:3D_ISDtypes} for the three 3-D simulations performed: constant mobility (blue triangles), dissimilar mobilities with phase separated initial condition (red squares), and dissimilar mobilities with random noise initial condition (yellow diamonds).
	}
	\label{fig:3D_type_ev}
\end{figure}

Figure \ref{fig:3D_type_ev} shows how the integrated probability over each of these ISD regions identified in Fig.\ \ref{fig:3D_ISDtypes} evolves during coarsening.
The constant-mobility case and the dissimilar-mobility case with the phase separated initial condition (dissimilar mobilities PS IC) are identical at $t=10^4$ because both are simulated with $M=1$ during the phase separation stage.
However, the latter quickly diverges from the constant mobility case, and the probabilities of both dissimilar-mobility cases converge over time to the same late-time values in all of the four types of interfaces.
As compared to the constant-mobility structure, the dissimilar-mobility structures have less interfacial area that is nearly flat or near the ISD peak (Figs.\ \ref{fig:3D_type_ev}a and \ref{fig:3D_type_ev}b), and significantly more interfacial area corresponding to high-mobility-phase necks (Fig.\ \ref{fig:3D_type_ev}d), along with slightly more low-mobility-phase necks (Fig.\ \ref{fig:3D_type_ev}c).

The constant-mobility ISD does not evolve substantially, except for a slight increase in the amount of area corresponding to necks at early times.
The neck probabilities must be statistically identical in the constant-mobility ISD due to kinetic and compositional symmetries, and therefore the observed changes in neck probabilities at late times represent statistical uncertainty due to our finite sample size.
In all cases, the morphologies quantified in Fig.\ \ref{fig:3D_type_ev} appear to have converged by $t>2\times 10^5$ ($t^{1/3}>58$).
This finding is confirmed in Appendix B, which examines convergence of the time-dependent ISDs to time-averaged ISDs representing the converged state.

\begin{figure}
	\centering
	\includegraphics[width = 14.4cm]{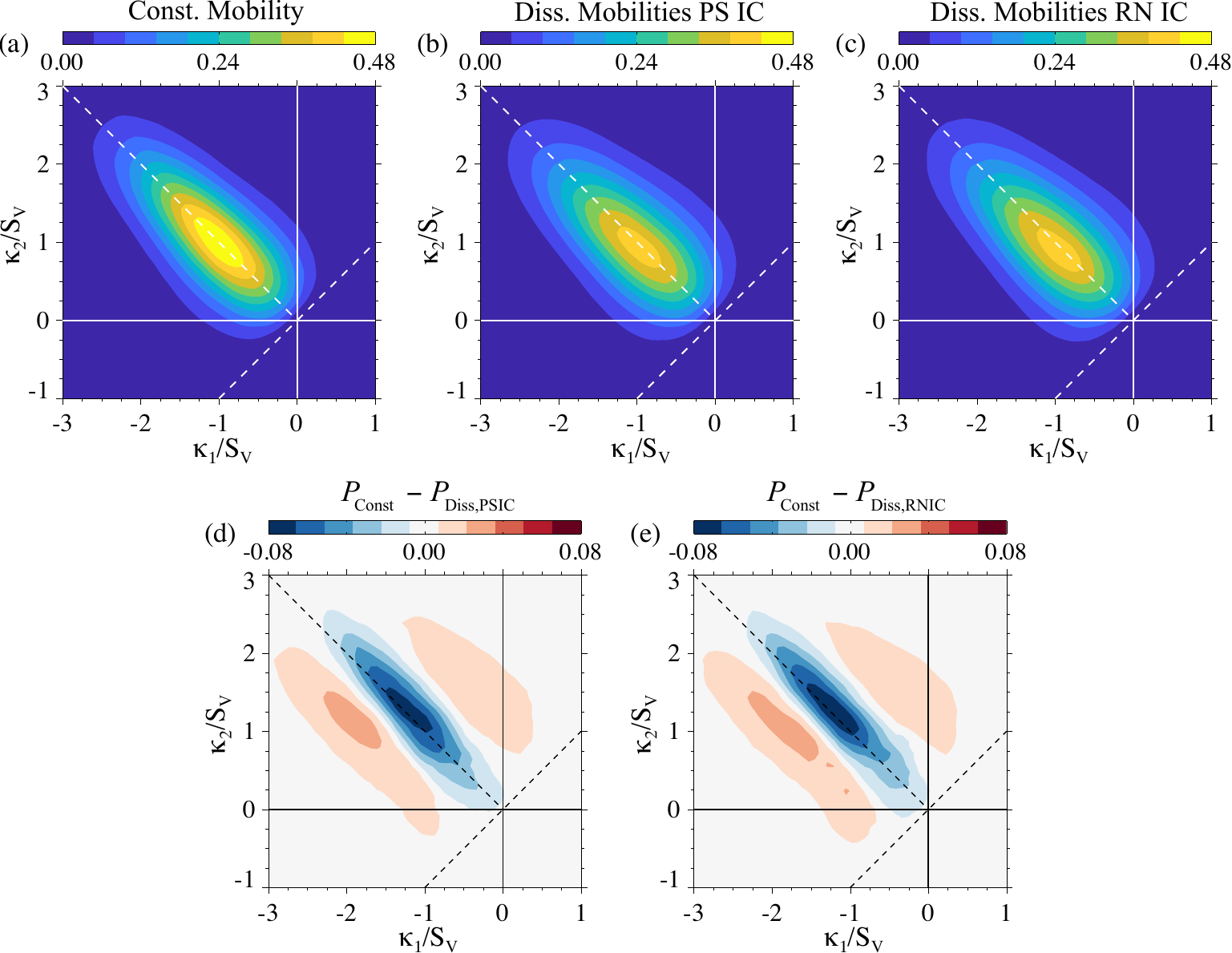}
	\caption{
		Late-time ISDs for (a) constant mobility, (b) dissimilar mobilities with phase separated initial condition, and (c) dissimilar mobilities with random noise initial condition, time-averaged over the interval $2\times 10^5< t \le 4\times 10^5$.  The differences between the constant-mobility ISD in (a) and the dissimilar-mobilities ISDs in (b) and (c) are shown in (d) and (e), respectively.
	}
	\label{fig:3D_ISD}
\end{figure}

Figures \ref{fig:3D_ISD}a-c show these converged ISDs, which are time-averaged over the interval $2\times 10^5< t \le 4\times 10^5$, for the constant-mobility, dissimilar-mobility PS IC, and dissimilar-mobility RN IC cases, respectively.
All three ISDs are concentrated around the line of zero mean curvature, $H=(\kappa_1+\kappa_2)/2=0$, which is indicated by a dashed line extending from the origin to the upper left corner.
The constant-mobility ISD (Fig.\ \ref{fig:3D_ISD}a) is symmetric about the $H=0$ line, as expected due to the symmetry between the phases at \%50 volume fraction.
However, as was inferred in Figs.\ \ref{fig:3D_type_ev}c and \ref{fig:3D_type_ev}d, the dissimilar-mobility ISDs in Figs.\ \ref{fig:3D_ISD}b and \ref{fig:3D_ISD}c are asymmetric, with more area corresponding to high-mobility-phase necks than to low-mobility-phase necks.
The dissimilar-mobility ISDs are also broader than the constant-mobility ISD, having a larger standard deviation of scaled mean curvature, $\sigma_{H/S_V} = 0.37$ vs.\ $\sigma_{H/S_V} = 0.32$.
The differences between the constant-mobility ISD (Fig.\ \ref{fig:3D_ISD}a) and the dissimilar-mobilities ISDs in Figs.\ \ref{fig:3D_ISD}b and \ref{fig:3D_ISD}c are shown in Figs.\ \ref{fig:3D_ISD}d and \ref{fig:3D_ISD}e, respectively.
These differences are quantitatively much larger than the difference between dissimilar-mobilities ISDs themselves: the integral of the absolute difference $|P_{Const}-P_{Diss.}|$ over the entire ISD is 0.135 for both dissimilar-mobilities cases, whereas it is only 0.019 between the dissimilar-mobility ISDs.
Since the differences between the dissimilar-mobilities ISDs are small, Figs.\ \ref{fig:3D_ISD}d and \ref{fig:3D_ISD}e are qualitatively similar: they both show decreases in area along the $H=0$ line and increases in area at higher and lower $H/S_V$, indicating the broadening of the peaks of the dissimilar-mobilities ISDs.
Figures \ref{fig:3D_ISD}d and \ref{fig:3D_ISD}e also show the asymmetry of the dissimilar mobilities ISDs.
In particular, the decrease in area near the $H=0$ line is larger for positive $H/S_V$, and away from the $H=0$ line there is a greater increase in area with negative $H/S_V$ than positive $H/S_V$.
While these asymmetries would be expected to decrease the average mean curvature $\left< H/S_V \right>$, they are offset by a long `tail’ of increased area with large positive values of $H/S_V$, resulting in $\left< H/S_V \right> = 0.00$.
Thus, despite the volume fraction of the two phases being essentially equal, the ISDs of the dissimilar mobilities cases are not symmetric about $H=0$, and they are skewed rather than shifted away from $H=0$.

This asymmetry in the dissimilar-mobility ISD may originate from the asymmetry in kinetics of coarsening of the two populations of necks.
Low-mobility-phase necks (which have positive mean curvature) are surrounded by high-mobility phase, and they disappear more rapidly than necks containing high-mobility phase that are surrounded by the low-mobility phase \cite{Aagesen2010,Aagesen2011}.
High-mobility-phase necks disappear more slowly, which may result in their increased prevalence in the structure compared to low-mobility-phase necks.
However, unlike high-mobility-phase particles in the 2-D case, high-mobility-phase necks can still evolve away through diffusion within the neck region.
Thus, the structure  does appear to reach a bicontinuous steady state, unlike the 2-D dissimilar-mobility case.

\subsection{Kinetics \label{sec:3Dkinetics}}
\begin{figure}
	\centering
	\includegraphics[width = 7.6cm]{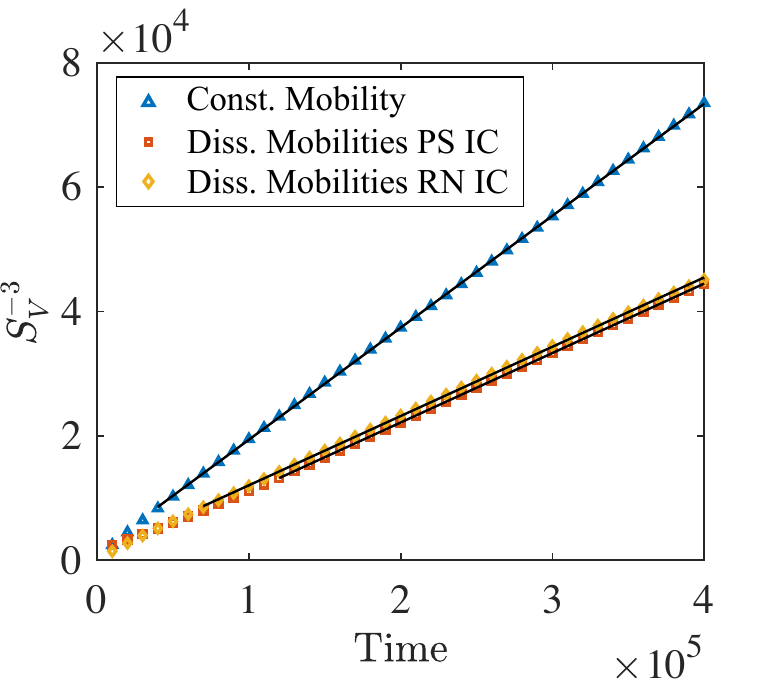}
	\caption{
		Coarsening kinetics for the 3-D structures.  The cube of the characteristic length, $S_V^{-3}$, is plotted vs.\ time for the three simulations: constant mobility (blue triangles), dissimilar mobilities with phase separated initial condition (red squares), and dissimilar mobilities with random noise initial condition (yellow diamonds).  Linear fits are shown as solid black lines.
	}
	\label{fig:3D_kinetics}
\end{figure}

Coarsening kinetics of the 3-D structures are illustrated in Fig.\ \ref{fig:3D_kinetics}, which shows the cube of characteristic length, $S_V^{-3}$, vs.\ time for each simulation alongside linear fits.
The dissimilar-mobility conditions have very similar coarsening rates, with equations of fit $S_V^{-3} = 0.112t - 191$ for PS IC and $S_V^{-3} = 0.111t + 905$ for RN IC, while the constant-mobility case coarsens more quickly, with $S_V^{-3} = 0.180t + 905$.
The coarsening rate constant for the constant-mobility case (the coefficient of $t$ in the fit) is in reasonable agreement with a literature value \cite{Kwon2010} of $0.173$ after rescaling to compensate for the different simulation parameters (see Appendix A).
All cases produce excellent fits, with $R^2 = 0.99998$ and  $R^2 = 0.99986$ for the dissimilar-mobility PS IC and RN IC cases, respectively, and  $R^2 = 0.99996$ for the constant mobility case.

Transient coarsening kinetics are observed in Fig.\ \ref{fig:3D_kinetics} prior to steady-state.
In the constant-mobility case, the rate of change of $S_V^{-3}$ (i.e., the instantaneous coarsening rate constant, $dS_V^{-3}/dt$) decreases as it approaches steady state.
The instantaneous coarsening rate constant of the dissimilar-mobility PS IC structure increases over time while that of the dissimlar-mobility RN IC structure decreases, resulting in a crossover of their $S_V^{-3}$ vs.\ time plots in Fig.\ \ref{fig:3D_kinetics} near $t=4\times 10^4$.
In the next section, we provide a model that may explain this difference in transient coarsening kinetics and the overall effect of dissimilar mobilities on kinetics.

\subsection{\label{sec:morphkin} Relationship between Kinetics and Morphology}
In this section, we examine a relationship between morphology and kinetics that we expect to be valid for the 3-D cases, namely that the coarsening rate constant $k=dS_V^{-3}/dt$ is proportional to the variance of scaled mean curvature, $\sigma_{H/S_V}^2$ and inversely proportional to a scaled diffusional interaction distance, $\hat \lambda$.
We derive this relationship by considering the local relationship between interfacial velocity and curvature and by employing statistical assumptions that are justified by previous analysis of the constant-mobility case \cite{Park2017}.
The details of this derivation are given in Appendix C, and only the assumptions and key results are presented here.

To begin, the interfacial velocity due to coarsening via bulk diffusion can be expressed as
\begin{equation}
\label{eq:velocity_full}
    v = \frac{\vec n \cdot M \nabla \mu |^+_-}{\phi_0 |^+_-},
\end{equation}
where the diffusive flux, $\vec j=-M\nabla \mu$, has been substituted into Eq.\ \ref{eq:interfacial_velocity}, and $|^+_-$ denotes the difference in the value of $\vec n \cdot \nabla \mu$ or $\phi$ between the high-mobility ($+$) and low-mobility ($-$) sides of the interface.
The gradient of chemical potential $\nabla \mu$ in each phase results from the solution to the Laplace equation (Eq.\ \ref{eq:laplace}) with boundary conditions set by the Gibbs-Thomson effect (Eq.\ \ref{eq:GT_mu}) at the interface.
To take advantage of the fact that $\mu$ is known at the interface, we express $\vec n \cdot \nabla \mu |^-$ and $\vec n \cdot \nabla \mu |^-$ (the normal components of $\nabla \mu$ at the high- and low-mobility sides of the interface, respectively) in terms of $H$ and two unknown variables each, interaction distances $\lambda^\pm$ and interaction mean curvatures $H_i^\pm$,
\begin{equation}
\label{eq:grad_mu}
    \vec n \cdot \nabla \mu |^+ =  \frac{2\gamma}{\left(\phi_0|_-^+\right)} \frac{H_i^+-H}{\lambda^+}, \;\;\;\; \vec n \cdot \nabla \mu |^- = - \frac{2\gamma}{\left(\phi_0|_-^+\right)} \frac{H_i^--H}{\lambda^-}.
\end{equation}
The complete expression for the interfacial velocity (combining Eqs.\ \ref{eq:velocity_full} and \ref{eq:grad_mu}) is then
\begin{equation}
\label{eq:velocity_full_subs}
    v = \frac{2\gamma}{\left(\phi_0|_-^+\right)} \left( M^+ \frac{H_i^+-H}{\lambda^+} + M^- \frac{H_i^--H}{\lambda^-} \right).
\end{equation}
$H_i^\pm$ and $\lambda^\pm$ provide enough degrees of freedom for Eq.\ \ref{eq:grad_mu} to represent all possible combinations of $H$ and $\vec n \cdot \nabla \mu |^\pm$ within the structure.
However, $H_i^\pm$ and $\lambda^\pm$ cannot be uniquely determined from Eq.\ \ref{eq:grad_mu} alone.
The interaction distances $\lambda^\pm$ must be positive, and therefore $H_i^\pm$ specify the signs of $\vec n \cdot \nabla \mu |^\pm$.
However, the magnitudes of $\vec n \cdot \nabla \mu |^\pm$ depend on both $\lambda^\pm$ and $H_i^\pm$.
These variables could be determined uniquely using additional information from the diffusion field \cite{DeHoff1991}, but this is not necessary for deriving information about the structure-wide dynamics.
Instead, we only need two statistical assumptions about $\lambda^\pm$ and $H_i^\pm$.
The first is that $\lambda^+$ and $\lambda^-$ are uncorrelated from $H_i^+$ and $H_i^-$, respectively, and from $H$.
This was proposed originally by DeHoff \cite{DeHoff1991}, but he incorrectly applied it to systems of spherical particles, where the solution to the Laplace equation in spherical coordinates dictates that $\lambda \propto 1/H$ \cite{Ratke2002,Fife2014}.
We will leave to future work the establishment of a rigorous geometric limit in which $\lambda$ is uncorrelated to $H$ and $H_i$.
The second assumption is that the averages of $H_i^+$ and $H_i^-$ over all interfaces with the same value of $H$ are always equal to $\left<H\right>$.
This can be interpreted as a mean-field approximation for average diffusional interactions, with the mean field set by $\left<H\right>$.
Mean-field approximations are widely used in models of both particulate structures (e.g., LSW theory \cite{Ratke2002}) and those proposed for more complex structures \cite{Fife2014,Park2017}.

With the assumptions stated above, the following intermediate result can be derived from Eq.\ \ref{eq:velocity_full_subs},
\begin{equation}
    \label{eq:vh_small}
    \left< v \right>_H= \frac{2\gamma }{\left(\phi_0|_-^+\right)^2} \left( M^+ \left< \frac{1}{\lambda^+}\right> + M^- \left< \frac{1}{\lambda^-}\right> \right) \left(\left< H \right> - H \right),
\end{equation}
where $\left<v\right>_H$ is the average interfacial velocity for all interfaces within the structure that have the same mean curvature.
Park et al.\ \cite{Park2017} computed $\left< v \right>_H$ and fit it to a cubic polynomial of $H$, with both $\left< v \right>_H$ and $H$ scaled by characteristic length and time.
At volume fractions of 50\% and 40\%/60\%, the quadratic and cubic terms of these fits are small for most of the interface, and can be neglected with minimal error, resulting in a linear relationship between $\left< v \right>_H$ and $H$ that matches Eq.\ \ref{eq:vh_small}.
Thus, the results of Park et al.\ \cite{Park2017} appear to validate the assumptions of our model, at least for the constant-mobility case.

To obtain the final relationship between $\sigma_{H/S_V}^2$ and $dS_V^{-3}/dt$, we employ the geometric identity
\begin{equation}
\label{eq:area_change}
  \frac{dA_T}{dt} = 2 \int_S H v dA,
\end{equation}
which relates the rate of change of total interfacial area $dA_T/dt$ to local interfacial mean curvature $H$ and velocity $v$ ($\int_S dA$ denotes integration over the entire interface).
The instantaneous coarsening rate constant is related to $dA_T/dt$ by
\begin{equation}
\label{eq:rate_dadt}
    \frac{dS_V^{-3}}{dt} = -3S_V^{-3} \frac{1}{A_T} \frac{dA_T}{dt}.
\end{equation}
The final expression, derived using Eqs.\ \ref{eq:velocity_full_subs}, \ref{eq:area_change}, and \ref{eq:rate_dadt} with our two assumptions, is
\begin{equation}
\label{eq:kin_morph_gen}
  \frac{dS_V^{-3}}{dt} = S_V^{-1} \frac{12\gamma}{\left(\phi_0|_-^+\right)^2} \left( M^+ \left< \frac{1}{\lambda^+}\right> + M^- \left< \frac{1}{\lambda^-}\right> \right)  \sigma_{H/S_V}^2,
\end{equation}
where $\left< 1/\lambda^+ \right>$ and $\left< 1/\lambda^- \right>$ are the area-weighted averages of $1/\lambda^+$ and $1/\lambda^-$, respectively, over the entire interface.
Details of the intermediate steps are provided in Appendix C.

Equation \ref{eq:kin_morph_gen} relates the coarsening rate constant to the scaled morphology with two unknowns, $\left< 1/\lambda^+ \right>$ and $\left< 1/\lambda^- \right>$.
For highly dissimilar mobilities, it may be appropriate to neglect the contribution of the low-mobility phase (the term containing $M^-$ in Eq.\ \ref{eq:kin_morph_gen}).
For constant mobility at 50\% volume fraction, it may be valid to assume that $\left< 1/\lambda^+ \right> = \left< 1/\lambda^- \right>$ based on the symmetry between the phases.
In general, however, $\left< \frac{1}{\lambda^+}\right>$ and $\left< \frac{1}{\lambda^-}\right>$ cannot be determined from the coarsening rate constant alone.
Therefore, we combine them into a single parameter, the scaled average interaction distance
\begin{equation}
\label{eq:lambda_hat}
  \hat \lambda = S_V \frac{2 \bar M}{M^+ \left< \frac{1}{\lambda^+}\right> + M^- \left< \frac{1}{\lambda^-}\right>},
\end{equation}
where the characteristic mobility of the structure $\bar M$ is the average based on a rule of mixtures, $\bar M = ( V^+ M^+ + V^- M^-)/ V$, in which $ V^+$ and $ V^-$ are the nominal volumes of high- and low-mobility phase, respectively, and $ V =  V^+ + V^-$.
Equation \ref{eq:kin_morph_gen} then becomes
\begin{equation}
\label{eq:kin_morph}
  \frac{dS_V^{-3}}{dt} =  \frac{24\gamma \bar M}{\left(\phi_0|_-^+\right)^2} \frac{ \sigma_{H/S_V}^2}{\hat \lambda},
\end{equation}
which relates the scaled morphology to the coarsening rate constant with a single unknown parameter, $\hat \lambda$, since the remaining parameters $\bar M$, $\gamma$, and $\left(\phi_0|_-^+\right)^2$ are known constants for a given simulation.
Thus, the instantaneous coarsening rate constant is directly proportional to $\sigma_{H/S_V}^2$ and inversely proprotional to $\hat \lambda$.
This relationship provides a framework for interpreting the differences in kinetics between the 3-D cases, and we test it against the morphological evolution observed in the dissimilar-mobilities cases.

\begin{figure}
	\centering
	\includegraphics[width = 7.4cm]{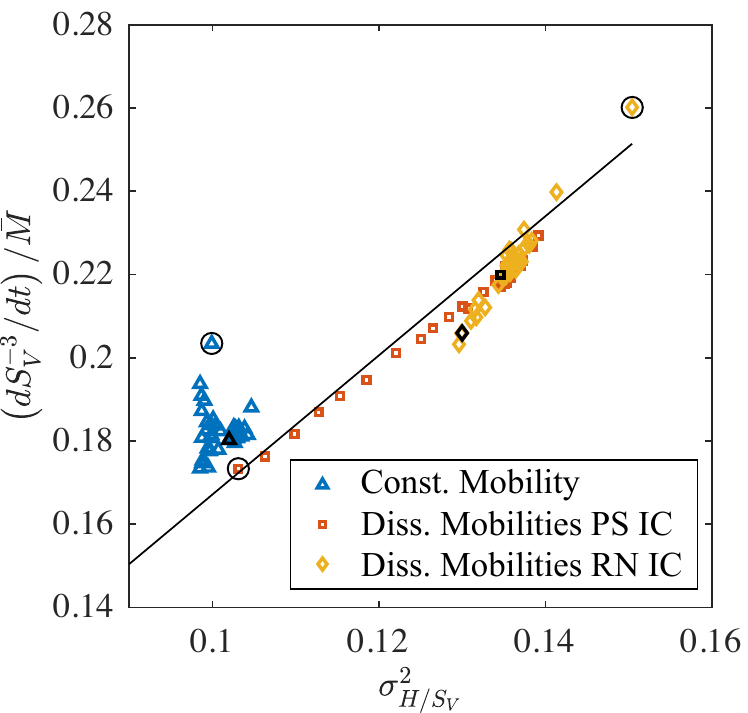}
	\caption{
		Plot of $(dS_V^{-3}/dt)/\bar M$, the instantaneous coarsening rate constant divided by the average mobility, vs.\ the variance of scaled mean curvature, $\sigma^2_{H/S_V}$, for the 3-D simulations: constant mobility (blue triangles), dissimilar mobilities with phase separated initial condition (red squares), and dissimilar mobilities with random noise initial condition (yellow diamonds).
		A fit of Eq.\ \ref{eq:kin_morph} to all three data sets is indicated by a solid black line.  The earliest data points (calculated between $t=10^4$ and $t=2\times 10^4$) are circled, and the final data points (calculated between $t=3.9\times 10^5$ and $t=4\times 10^5$) are noted by black symbols.
	}
	\label{fig:3D_dehoff}
\end{figure}

To examine the relationship presented in Eq.\ \ref{eq:kin_morph}, $(dS_V^{-3}/dt)/\bar M$ is plotted against $\sigma_{H/S_V}^2$ in Fig.\ \ref{fig:3D_dehoff} for the three 3-D simulations.
The coarsening rate constants are calculated from the data in Fig.\ \ref{fig:3D_kinetics} by a centered difference between output steps, i.e., $k(t_{n+1/2}) = [S_V^{-3}(t_{n+1}) - S_V^{-3}(t_n)]/(t_{n+1}-t_{n})$, where $t_n$ is the time of the $n$th output step.
The variance in scaled mean curvature $\sigma_{H/S_V}^2$ was calculated at $t_{n+1/2}$ by a simple average.
The earliest data points (calculated between $t=10^4$ and $t=2\times 10^4$) are circled, and the final data points (calculated between $t=3.9\times 10^5$ and $t=4\times 10^5$) are noted by black symbols.
Figure \ref{fig:3D_dehoff} thus includes both transient and steady-state regimes, where the clusters of points near the black symbols correspond to the steady-state regime.
Equation \ref{eq:kin_morph} is valid during the transient regime (provided its original assumptions hold), but both $\hat \lambda$ and $\sigma_{H/S_V}^2$ may vary.
A single fit to Eq.\ \ref{eq:kin_morph} using all of the data resulted in $\hat \lambda = 0.48$, and is shown as a solid black line.

It should be noted that $\sigma_{H/S_V}^2$ does not evolve substantially in the constant-mobility structure, yet the coarsening rate constant for this case does evolve in Fig.\ \ref{fig:3D_dehoff}, at least at early times.
The coarsening rate constant decreases rapidly from its initial value (the circled blue triangle in Fig.\ \ref{fig:3D_dehoff}) to its steady state value, indicated by the cluster of points around the black triangle.
In our proposed model (Eq.\ \ref{eq:kin_morph}), this decrease in $dS_V^{-3}/dt$ corresponds to an increase in the scaled interaction distance $\hat \lambda$.
As an approximate measure of interaction distance, increasing $\hat \lambda$ suggests longer or more tortuous paths through the structure for diffusive fluxes.
However, since we do not evaluate $\hat \lambda$ directly, we cannot rule out a breakdown in our assumptions within this regime (i.e., $\lambda^\pm$ may in fact correlated to $H$ and $H_i^\pm$ at early times).
The fact that the change in $dS_V^{-3}/dt$ in the constant-mobility case occurs early in the simulation and without any obvious change in the morphology (see Fig.\ \ref{fig:3D_type_ev} and Appendix B) raises the possibility that the concentration field is still converging to steady state.
This convergence would explain an increase in $\hat \lambda$ that matches the initial decrease in $dS_V^{-3}/dt$.
Such an effect may also be present in the dissimilar-mobilities simulations at early times.

The most compelling case for the relationship between $dS_V^{-3}/dt$ and $\sigma_{H/S_V}^2$ is provided by the dissimilar-mobility PS IC case, in which coarsening with dissimilar mobilities was initialized from the constant-mobility phase separation morphology.
The simulation data for this case begins close to the cluster of points from the constant-mobility case, where the structure has smaller $\sigma_{H/S_V}^2$ and $dS_V^{-3}/dt$ than its final state.
As $\sigma_{H/S_V}^2$ increases, the rate constant increases, eventually reaching the cluster of values representing the steady state of the RN IC simulation.
The interaction distance $\hat \lambda$ undergoes relatively little evolution, resulting in a nearly linear relationship between  $\sigma_{H/S_V}^2$ and the coarsening rate constant.
The evolution of $dS_V^{-3}/dt$ and $\sigma_{H/S_V}^2$ in the PS IC case takes place over a relatively long time interval (more than ten output steps/data points), ruling out more rapid transient dynamics as a source of the change in $dS_V^{-3}/dt$.
The relationship between $\sigma_{H/S_V}^2$ and the rate constant is also observed to the RN IC case, where the structure resulting from the random noise initial condition has a wider distribution in $H$ than the self-similar structure (see the circled yellow diamond in Fig.\ \ref{fig:3D_dehoff}).
Evolution is rapid in this case, with most of the change in $\sigma_{H/S_V}^2$ occurring between the first and third data points.
When Eq.\ \ref{eq:kin_morph} is fitted only to the dissimilar-mobility datasets we obtain $\hat \lambda = 0.49$, while when it is fitted only to the constant-mobility dataset, $\hat \lambda$ is $0.44$.

We can now explain the effect of dissimilar mobilities on kinetics observed in Section \ref{sec:3Dkinetics}, where the steady-state rate constants of the dissimilar-mobilities cases are approximately $0.6$ times that of the constant-mobility case ($0.11$ and $0.18$, respectively).
Of the parameters in Eq.\ \ref{eq:kin_morph}, the largest change due to dissimilar mobilities is the decrease in $\bar M$ by $\sim$50\%, followed by the increase in $\sigma_{H/S_V}^2$ by $\sim$30\%.
The calculated change in $\hat \lambda$ is relatively smaller; it increases by $\sim$10\% for the dissimilar-mobilities cases.
Combining all of these effects results in the observed difference in rate constant: $k_{Diss}=(0.5 \times 1.3 / 1.1) k_{Const}=0.6k_{Const}$.

\subsection{\label{sec:3D_diss} Discussion}
Overall, we have found that 3-D structures with dissimilar mobilities at the critical composition (50\% volume fraction) coarsen self-similarly with the expected $t^{1/3}$ power law.
This provides a dramatic contrast with the 2-D case, where an initially complex structure breaks down into a system of high-mobility particles in a low-mobility matrix.
The key difference between these cases is the continuity of the high-mobility phase.
Continuity of the high-mobility phase in 3D ensures that the entire structure is coarsening at a rate determined by the high-mobility phase, as opposed to the 2-D case where there two types of features evolving at different rates.
In 2D, a structure cannot be bicontinuous, and the complex morphologies observed during coarsening with constant mobility are only possible due to the symmetry between the phases at 50\% volume fraction, which prevents them from differentiating into a particle phase and a matrix phase.
Even at volume fractions of 45\%/55\% with constant mobility, this symmetry is broken and a particle/matrix structure is observed \cite{Rogers1989}.
Dissimilar mobilities also break the symmetry between phases, leading to the morphological transition we describe in Section \ref{sec:2D_diss}.
In 3D, bicontinuous structure are stable during coarsening via bulk diffusion to volume fractions at or below 36\% \cite{Kwon2010}.
Thus, it should not be surprising that the structure at 50\% is still bicontinuous with dissimilar mobilities.

Examining the combined effects of volume fraction and dissimilar mobilities is an interesting topic for future work.
In 2D, it is possible that a more complex structure would be recovered at higher volume fractions of high-mobility phase due to more rapid coalescence of high-mobility regions.
In 3D, analogy could be drawn from recent studies of hydrodynamic coarsening with dissimilar viscosities \cite{Bouttes2016,Henry2019}.
These studies find that structures where the minority phase has low viscosity break up into particles at higher volume fractions than structures where the minority phase has high viscosity.
The explanation provided for this symmetry-breaking is that necks/ligaments containing low-viscosity phase pinch off more quickly than necks/ligaments containing high-viscosity phase \cite{Bouttes2016}.
In our system, necks containing low-mobility phase pinch off more quickly than necks containing high-mobility phase \cite{Aagesen2010,Aagesen2011}.
By analogy, we would expect structures containing high-mobility phase to be stable at lower volume fractions than structures containing low-mobility phase.

While Ref.\ \cite{Bouttes2016} provides an interesting perspective on the relationship between system parameters (material properties and volume fraction) and coarsening morphology, our main theoretical contribution is the relationship between morphology and kinetics in Eq.\ \ref{eq:kin_morph_gen}.
This relationship between $\sigma_{H/S_V}^2$ and $dS_V^{-3}/dt$ is likely to be applicable to complex microstructures observed in experimental systems.
Fife et al.\ \cite{Fife2014} observed correlations between $H$ and $\left<v\right>_H$ during coarsening of an Al-Cu dendritic solid-liquid mixture.
However, coarsening of dendritic structures is also affected by the spatial distribution of dendrites \cite{Sun2017} and the presence of multiple length scales associated with primary, secondary, and tertiary arms \cite{Sun2018}.
Additionally, our simulations ignore the effect of crystallinity, which would prevent different solid crystals from coalescing without forming a grain boundary.
Future work may explore the effect of these effects on coarsening dynamics, as well as those of volume fraction and the ratio of mobilities.

\section{Conclusions}
Coarsening of a two-phase system in which the phases had dissimilar mobilities was studied at $50\%$ volume fraction.
Simulations were conducted in two and three dimensions using the Cahn-Hilliard model with a concentration-dependent mobility formulated to reduce the effect of small deviations in concentration, such as those due to the Gibbs-Thomson effect.
Simulations with constant mobility were conducted for comparison.
The simulated morphologies were characterized by the characteristic length, statistics of mean curvature, and the interfacial shape distribution.
Quantitative analysis of the interfacial shape distribution was applied to the 3-D simulation results.

A morphological transition during coarsening was identified in two dimensions when the mobility was dissimilar that was not observed when the mobility was constant.
In this transition, an initially layered structure transforms into a system of high-mobility-phase particles embedded in a low-mobility matrix.
This morphological transition resulted in a decrease of the coarsening rate constant over time, which explains why previous studies \cite{Sheng2010,Ju2015} did not find agreement with the $L \propto t^{1/3}$ power law but rather suggested smaller coarsening exponents.
Morphological evolution was also observed in the 2-D constant-mobility case, and its kinetics agreed with the $t^{1/3}$ power law after an initial transient stage.

The 3-D simulations resulted in bicontinuous morphologies that evolved self-similarly.
Close agreement was found between the late-time morphologies of two dissimilar-mobility simulations that differed only in their initial conditions.
The self-similar morphology for the dissimilar-mobility cases has greater variance in scaled mean curvature than the constant-mobility morphology.
The dissimilar-mobility morphology also possesses slight asymmetry caused by the asymmetry in mobility: it contains more area corresponding to necks surrounded by low-mobility phase, which evolve more slowly, than area corresponding to necks surrounded by high-mobility phase, which evolve more quickly.

Based on theory and simulations, the primary difference in the kinetics of coarsening between the dissimilar-mobility and constant mobility systems were determined to be a factor of two smaller kinetic coefficient in the dissimilar-mobility case due to the lack of diffusion in one of the phases.
The coarsening kinetics of the 3-D cases agreed well with the theoretical $t^{1/3}$ power law after initial transient stages.
During the transient stages of the dissimilar-mobilities simulations, a nearly-linear relationship was observed between the variance in scaled mean curvature, $\sigma_{H/S_V}^2$, and the instantaneous coarsening rate constant, $k = dS_V^{-3}/dt$, calculated between simulation output steps.
We have derived this relationship using statistical assumptions and shown that it is consistent with previous analysis of interfacial velocities for the constant-mobility case \cite{Park2017}.
Our derivation contains a single free parameter, the scaled average interaction distance $\hat \lambda$, for which the constant- and dissimilar-mobility cases yielded similar values.
The resulting expression for the instantaneous kinetic coefficient provides fundamental insight into coarsening phenomena within complex microstructures; after further validation, it may prove to be a practical analytical model for materials engineering.

\setcounter{equation}{0}
\renewcommand\theequation{A.\arabic{equation}}
\section*{Appendix A: Nondimensionalization and Rescaling}
Solutions $\phi(\vec x, t)$ to Eqs.\ \ref{eq:CH} \& \ref{eq:CH_mu} with different parameters correspond to solutions to a dimensionless Cahn-Hilliard equation scaled in length and time.
To make comparisons to other works that use different forms of the Cahn-Hilliard equation and associated parameters (i.e., \cite{Kwon2010} and \cite{Dai2016}), here we provide a procedure for rescaling to match our parameters.
In addition to the parameters $\epsilon$, $M$, and $W$, the primary difference is in the bulk free energy $f(\phi)$ that affects the range of $\phi$ between the two equilibrium values.
We consider only models with the same functional forms of the bulk free energy that can be rescaled to a dimensionless function $g(\tilde \phi)$ by
\begin{equation}
    g(\tilde \phi) = 16  \left(\frac{1}{4} - \tilde \phi^2 \right)^2 = \frac{1}{\hat F} f\left(\frac{\phi - \phi_c}{\hat \Phi}\right),
\end{equation} 
where $\hat F$ is the height of the double well free energy that is used as the energy density scale,  $\hat \Phi$ is the difference between the two equilibrium concentrations, and $\phi_c$ is the midpoint of the equilibrium concentrations.
The concentration $\phi_c$ only shifts the concentration range and will not affect spatial or temporal scaling.

While any scaling length can be chosen for nondimensionalization, we select one that is associated with the interfacial width to rescale all models.
This scaling length $\hat L$ is defined as 
\begin{equation}
\label{eq:char_l}
    \hat L = \frac{\epsilon \hat \Phi}{\sqrt{2\hat F}}.
\end{equation}
The characteristic time $\hat T$ associated with $\hat L$ is then determined by dimensional analysis of Eqs.\ \ref{eq:CH} \& \ref{eq:CH_mu},
\begin{linenomath}
\begin{equation*}
    \frac{\hat \Phi}{\hat T} = \frac{M}{\hat L^2} \frac{\hat F}{\hat \Phi}
\end{equation*}
\end{linenomath}
\begin{equation}
\label{eq:char_t}
    \hat T =  \frac{\hat \Phi^2 \hat L^2}{M\hat F}
\end{equation}
Defining the dimensionless coordinates $\tilde t = t/\hat T$, $\tilde x = x/\hat L$, and the dimensionless concentration $\tilde \phi = (\phi - \phi_c)/\hat \Phi$, the non-dimensional Cahn-Hilliard equation is now given by
\begin{equation}
    \frac{\partial \tilde \phi}{\partial \tilde t} = \tilde \nabla^2 \left[ g'(\tilde \phi) - 2 \tilde \nabla^2 \tilde \phi \right].
\end{equation}

For two given simulations $A$ and $B$, we can now rescale the simulation time $t_A$ from simulation $A$ to that corresponding to the time $t_B$ as measured in simulation $B$:
\begin{linenomath}
\begin{equation*}
    \tilde t = \frac{t_A}{\hat T_A} = \frac{t_B}{\hat T_B}
\end{equation*} 
\end{linenomath}
\begin{equation}
\label{eq:scale_t}
    t_B = \left( \frac{\hat T_B}{\hat T_A} \right) t_A.
\end{equation}
To compare the coarsening rate constant $k$ in Eq.\ \ref{eq:powerlaw} between simulations, we define the dimensionless coarsening rate constant $\tilde k = k \hat T/\hat L^3$.
The coarsening rate constant $k_A$ measured in simulation $A$ can now be rescaled to match the length and time scales employed in simulation $B$:
\begin{linenomath}
\begin{equation*} 
    \tilde k = \frac{k_A \hat T_A}{\hat L_A^3} = \frac{k_B \hat T_B}{\hat L_B^3}
\end{equation*} 
\end{linenomath}
\begin{equation}
\label{eq:scale_k}
    k_B = \left( \frac{\hat T_A \hat L_B^3}{ \hat T_B \hat L_A^3} \right) k_A.
\end{equation}
Eqs.\ \ref{eq:scale_t} and \ref{eq:scale_k} were used to compare the timescale in Ref.\ \cite{Dai2016} and the coarsening rate constant in Ref.\ \cite{Kwon2010}, respectively, to the corresponding values in the present work.

For example, Ref.\ \cite{Dai2016} uses $f(\phi)=\frac{1}{4}(\phi^2-1)^2$, $\epsilon=0.05$, and $M=1$.
This results in $\hat \Phi = 2$ and $\hat F = 0.25$, and, based on Eqs.\ \ref{eq:char_l} and \ref{eq:char_t}, the characteristic length of their system is $\hat L = 0.1414$ and the characteristic time is $\hat T = 0.32$.
In contrast, the present work uses $f(\phi)=\frac{W}{4}\phi^2 (1-\phi)^2$ (with $W=0.4$), $\epsilon = \sqrt{0.2}$, and $M=1$, which results in $\hat \Phi = 1$ and $\hat F = W/64 = 0.00625$.
Thus, our characteristic length is $\hat L = 4$ and time is $\hat T = 2560$.
In Ref.\ \cite{Dai2016}, the kinetics of a 2-D simulation with a one-sided mobility are fitted with times up to $t=8$.
Applying Eq.\ \ref{eq:scale_t}, this corresponds to
\begin{linenomath}
\begin{equation*}
    t = \frac{2560}{0.32} \times 8 = 6.4 \times 10^4
\end{equation*}
\end{linenomath}
with our parameters.

\setcounter{figure}{0}
\renewcommand\thefigure{B.\arabic{figure}}
\section*{Appendix B: Convergence to Late-Time Average ISDs}

\begin{figure}
	\centering
	\includegraphics[width = 9cm]{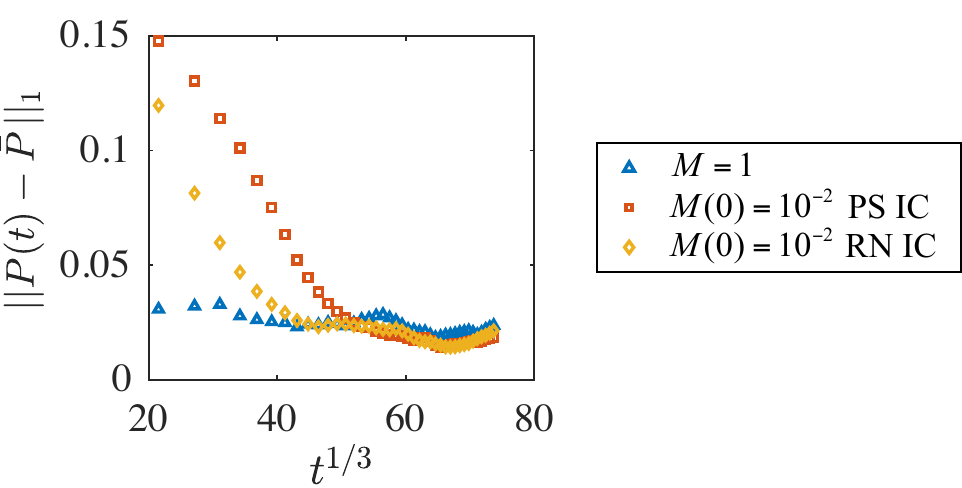}
	\caption{
		Plot of  $||P(t)-\bar P||_1$ vs.\ $t^{1/3}$ illustrating convergence of ISDs for the 3-D cases: constant mobility (blue triangles), dissimilar mobilities with phase separated initial condition (red squares), and dissimilar mobilities with random noise initial condition (yellow diamonds).
	}
	\label{fig:sup_convergence}
\end{figure}

Convergence of the time-dependent ISDs, denoted by $P(t)$ to the average ISDs, denoted by $\bar P$, is depicted in Fig.\ \ref{fig:sup_convergence}, which plots the $L^1$ norm of the difference $P(t)-\bar P$ vs.\ time.
This quantity, $||P(t)-\bar P||_1$, is the integral of the absolute difference $|P(t)-\bar P|$  over the domain of the ISD.
Differences between ISDs are sensitive to ISD resolution, which was $\Delta \kappa_1/S_V = \Delta \kappa_2/S_V = 0.08$ in this paper.
Since the integrals of the ISDs themselves are unity (because of the normalization in Eq.\ \ref{eq:ISD3D}), this measure of difference requires no additional normalization.
In all cases shown in Fig.\ S1, $||P(t)-\bar P||_1$ has converged to small, stable values ($0.02-0.03$) before $t^{1/3} \approx 58$, the beginning of the interval over which $\bar P$ is averaged, indicating that the average ISDs are representative of the converged, self-similar morphologies.

These values of $||P(t)-\bar P||_1$ during the converged regime provide an estimate of the uncertainty of our ISDs, which can be used to assess the statistical significance of the differences between ISDs for different conditions.
The $L^1$-normed difference between dissimilar-mobility ISDs is $0.019$, and the $L^1$-normed difference between either of them and the constant-mobility ISD is $0.135$.
The time-averaged dissimilar-mobility structures are therefore statistically indistinguishable, and they differ significantly from the constant-mobility structure.
The values of $||P(t)-\bar P||_1$ during the converged regime can also be used to assess when the structure has converged.
In Fig.\ \ref{fig:3D_kinetics}, kinetics were fitted over a timescale determined by the ad hoc criterion $||P(t)-\bar P||_1 \le 0.030$.

\setcounter{figure}{0}
\renewcommand\thefigure{C.\arabic{figure}}
\setcounter{equation}{0}
\renewcommand\theequation{C.\arabic{equation}}
\section*{Appendix C: Derivation of Relationship between Kinetics and Morphology}
In this appendix, we provide the intermediate steps in the derivation of the relationship between kinetics and morphology (specifically the coarsening rate constant $dS_V^{-3}/dt$ and the variance in scaled mean curvature $\sigma_{H/S_V}^2$) that is expressed in Eq.\ \ref{eq:kin_morph_gen}.
Our starting point is Eq.\ \ref{eq:velocity_full_subs}, which we restate here for convenience,
\begin{equation}
\label{eq:velocity_full_subs_appc}
    v = \frac{2\gamma}{\left(\phi_0|_-^+\right)} \left( M^+ \frac{H_i^+-H}{\lambda^+} + M^- \frac{H_i^--H}{\lambda^-} \right).
\end{equation}
To proceed from Eq.\ \ref{eq:velocity_full_subs_appc}, we must relate interfacial velocity to coarsening kinetics.
The rate of change of interfacial area $dA_T/dt$ is related to interfacial velocity and curvature by Eq.\ \ref{eq:area_change},
\begin{equation}
\label{eq:area_change_appc}
  \frac{dA_T}{dt} = 2 \int_S H v dA.
\end{equation}
The right-hand side of this equation represents an area-weighted integral of $v$ and $H$ over the entire structure.
The same quantity can be obtained by integrating in $v$ and $H$ over their ranges in the structure with an appropriate weight: the probability density function $f(v,H)$ for points having values of $v$ and $H$.
We therefore define the probability distribution function $f$ by
\begin{equation}
\label{eq:Hvdist_def}
f=\frac{1}{A_T}\frac{\partial^2 F_A^{Hv}}{\partial H \partial v},
\end{equation}
where, similarly to $F_A$ used to define the ISD in Eq.\ \ref{eq:ISD3D}, $F_A^{Hv}$ is a cumulative area distribution function representing the total area of interface with velocity less than or equal to $v$ and mean curvature less than or equal to $H$.
Now Eq.\ \ref{eq:area_change_appc} can be re-expressed as
\begin{equation}
\label{eq:area_change_int}
  \frac{dA_T}{dt} = 2 A_T \int \limits_{-\infty}^{\infty} \int \limits_{-\infty}^{\infty} H v f dv dH,
\end{equation}
and we can integrate in $v$, yielding
\begin{equation}
\label{eq:area_change_vh}
  \frac{dA_T}{dt} = 2 \int \limits_{-\infty}^{\infty} H \left< v \right>_H \frac{\partial F_A^{Hv}}{\partial H} dH,
\end{equation}
where $\left< v \right>_H$ is the average velocity at a given mean curvature,
\begin{equation}
\label{eq:vh_def}
    \left< v \right>_H = \frac{\int_{-\infty}^{\infty} v f dv}{\int_{-\infty}^{\infty} f dv},
\end{equation}
and from Eq.\ \ref{eq:Hvdist_def}, $\int_{-\infty}^{\infty} f dv=\partial F_A^{Hv}/\partial H$.
To derive the expression for $\left<v\right>_H$ (Eq.\ \ref{eq:vh_small}) that we compare to the results of Park et al.\ \cite{Park2017}, we first define a probability distribution function for the five variables from Eq.\ \ref{eq:velocity_full_subs_appc}, $H$, $\lambda^+$, $\lambda^-$, $H_i^+$, and $H_i^-$,
\begin{equation}
    f^* = \frac{1}{A_T} \frac{\partial^5 F_A^*(H,\lambda^+, \lambda^-, H_i^+, H_i^-)}{\partial H \partial \lambda^+ \partial \lambda^- \partial H_i^+ \partial H_i^- },
\end{equation}
where $F_A^*$ is the cumulative area distribution function for all five variables.
Since $f^*$ is essentially decomposing the variability of $f$ in $v$ at constant $H$ into four additional variables, we have $\partial F_A^*/\partial H = \partial F_A^{Hv}/\partial H$, and Eq.\ \ref{eq:vh_def} can be rewritten in terms of $f^*$ as
\begin{equation}
\label{eq:vh_def2}
    \left< v \right>_H = \frac{ \int_{-\infty}^{\infty} \int_{-\infty}^{\infty} \int_{0}^{\infty} \int_{0}^{\infty} v f^* d\lambda^+ d\lambda^- dH_i^+ dH_i^- }{ \int_{-\infty}^{\infty} \int_{-\infty}^{\infty} \int_{0}^{\infty} \int_{0}^{\infty} f^* d\lambda^+ d\lambda^- dH_i^+ dH_i^- } ,
\end{equation}
where the denominator of the right-hand side evaluates to $\partial F_A^*/\partial H$.
Now, multiplying both sides of Eq.\ \ref{eq:velocity_full_subs_appc} by $f^*$ and integrating in $H_i^\pm$ and $\lambda^\pm$ yields
\begin{equation}
\label{eq:vh_big}
       \left< v \right>_H \frac{\partial F_A^*}{\partial H} = \frac{2\gamma }{\left(\phi_0|_-^+\right)^2} \int \limits_{-\infty}^{\infty} \int \limits_{-\infty}^{\infty} \int \limits_{0}^{\infty} \int \limits_{0}^{\infty} \left(  M^+ \frac{H_i^+-H}{\lambda^+} + M^- \frac{H_i^--H}{\lambda^-}  \right) f^* d\lambda^+ d\lambda^- dH_i^+ dH_i^-,
\end{equation}
We now make the first of the assumptions cited in the main text: that $\lambda^\pm$ are uncorrelated with $H$ and $H_i^\pm$, i.e., that $f^* = f^H f^\lambda$, where
\[  f^H = \frac{1}{A_T}\frac{\partial^3 F_A^H(H,H_i^+,H_i^-)}{\partial H \partial H_i^+ \partial H_i^-}\;\;\; \mathrm{and}\;\;\; f^\lambda = \frac{1}{A_T} \frac{\partial^2 F_A^\lambda (\lambda^+,\lambda^-)}{ \partial \lambda^+ \partial \lambda^-}. \]
With this assumption, Eq.\ \ref{eq:vh_big} becomes
\begin{equation}
\label{eq:vh_medium}
     \left< v \right>_H \frac{\partial F_A^*}{\partial H} = \frac{2\gamma }{\left(\phi_0|_-^+\right)^2} \int \limits_{-\infty}^{\infty} \int \limits_{-\infty}^{\infty} \left[  M^+ \left<\frac{1}{\lambda^+} \right> (H_i^+-H) + M^- \left<\frac{1}{\lambda^-} \right> (H_i^--H)  \right] f^H dH_i^+ dH_i^-,
\end{equation}
where $\left< \frac{1}{\lambda^\pm} \right> = \int_0^\infty \int_0^\infty \frac{1}{\lambda^\pm} f^\lambda d\lambda^+ d\lambda^- = \frac{1}{A_T} \int_{S} \frac{1}{\lambda^\pm} dA$.
We now make the second assumption cited in the main text: that the averages of $H_i^+$ and $H_i^-$ over all interfaces with the same value of $H$ are equal to $\left<H\right>$ for all $H$, i.e., that
\begin{equation}
\label{eq:meanfield}
    \int \limits_{-\infty}^{\infty} \int \limits_{-\infty}^{\infty} H_i^+ f^H dH_i^+ dH_i^- = \int \limits_{-\infty}^{\infty} \int \limits_{-\infty}^{\infty} H_i^- f^H dH_i^+ dH_i^- = \left< H \right> \frac{\partial F_A^*}{\partial H}.
\end{equation}
This can be interpreted as a mean-field approximation for average diffusional interactions, with the mean field set by $\left<H\right>$.
With this assumption, Eq.\ \ref{eq:vh_medium} reduces to Eq.\ \ref{eq:vh_small}, which specifies a linear relationship between $\left< v \right>_H$ and $H$
\begin{equation}
    \label{eq:vh_small_appC}
    \left< v \right>_H= \frac{2\gamma }{\left(\phi_0|_-^+\right)^2} \left( M^+ \left< \frac{1}{\lambda^+}\right> + M^- \left< \frac{1}{\lambda^-}\right> \right) \left(\left< H \right> - H \right).
\end{equation}
Integrating Eq.\ \ref{eq:vh_small_appC} in $H$ and applying the mean-field-type assumption results in
\begin{equation}
    \int_{-\infty}^\infty \left< v \right>_H (\partial F_A^*/\partial H) dH = 0.
\end{equation}
The LHS of this equation is equivalent to $\int_S v dA$, which is proportional to the rate of change of volume fraction \cite{DeHoff1991}.
Thus, the assumption in Eq.\ \ref{eq:meanfield} is equivalent to assuming constant volume fraction if our primary assumption (that $\lambda^\pm$ are uncorrelated with $H$ and $H_i^\pm$) is valid.

We now use Eq.\ \ref{eq:vh_small_appC} to obtain the relationship between morphology and coarsening kinetics.
Substituting Eq.\ \ref{eq:vh_small_appC} into Eq.\ \ref{eq:area_change_vh} results in
\begin{equation}
\label{eq:area_change_final}
  \frac{dA_T}{dt} = - 2 A_T \frac{2\gamma }{\left(\phi_0|_-^+\right)^2} \left( M^+ \left< \frac{1}{\lambda^+}\right> + M^- \left< \frac{1}{\lambda^-}\right> \right) \sigma^2_H,
\end{equation}
where $\sigma^2_H = \left< H^2 \right> - \left< H \right>^2$ is the variance of mean curvature.
Since $S_V^{-1} = V/A_T$, where $V$ is the total volume of the domain, the coarsening rate constant $k = dS_V^{-3}/dt$ is related to $dA_T/dt$ by Eq.\ \ref{eq:rate_dadt},
\begin{equation}
\label{eq:rate_dadt_appC}
    \frac{dS_V^{-3}}{dt} = -3S_V^{-3} \frac{1}{A_T} \frac{dA_T}{dt}.
\end{equation}
Substituting Eq.\ \ref{eq:area_change_final} into Eq.\ \ref{eq:rate_dadt_appC} yields Eq.\ \ref{eq:kin_morph_gen},
\begin{equation}
\label{eq:kin_morph_gen_appC}
  \frac{dS_V^{-3}}{dt} = S_V^{-1} \frac{12\gamma}{\left(\phi_0|_-^+\right)^2} \left( M^+ \left< \frac{1}{\lambda^+}\right> + M^- \left< \frac{1}{\lambda^-}\right> \right)  \sigma_{H/S_V}^2,
\end{equation}
which relates the coarsening rate constant to the scaled morphology with two unknown parameters, $\left< 1/\lambda^+ \right>$ and $\left< 1/\lambda^- \right>$.

\section*{Acknowledgements} 
The authors gratefully acknowledge the generous support from the U.S.\ Department of Energy's Office of Science under Grants No.\ DE-SC0015394 and No.\ DE-FG02-99ER45782.
The simulations used computational resources provided by the Extreme Science and Engineering Discovery Environment (XSEDE), which is supported by National Science Foundation grant number OCI-1053575, under allocation No.\ TG-DMR110007, as well as the University of Michigan Advanced Research Computing.

\section*{Competing interests}
The authors declare no competing interests.

\section*{Data availability}
Raw and processed data required to reproduce these findings are available to download from the following links: 

\begin{itemize}
  \item www.doi.org/10.13011/m3-f11w-b549 (2D constant mobility dataset)
  \item www.doi.org/10.13011/m3-tha1-0v51 (2D dissimilar mobilities dataset)
  \item www.doi.org/10.13011/m3-w4pw-sy79 (3D constant mobility dataset)
  \item www.doi.org/10.13011/m3-4ybd-6e71 (3D dissimilar mobilities datasets)
\end{itemize}

\bibliographystyle{elsarticle-num}
\bibliography{biblio}

\end{document}